\documentclass[prd,preprint,tightenlines,floatfix,preprintnumbers,nofootinbib,eqsecnum]{revtex4}
 \usepackage{color}

\usepackage[dvips,final]{graphicx}
 \graphicspath{{./figs-ab/}}
  \usepackage{amssymb,amsmath}
\usepackage[cp1251]{inputenc}
 \usepackage[english,russian]{babel}

\newcommand{\va}[1]{\langle{#1}\rangle}                   
\newcommand{\gev}[1]{\relax\ifmmode{\text{ГэВ}^{#1}}      
                            \else{ГэВ$^{#1}${ }}\fi}      
\def\muF{\relax\ifmmode\mu_\text{F}^2\else{$\mu_\text{F}^2${ }}\fi}
\def\muR{\relax\ifmmode\mu_\text{R}^2\else{$\mu_\text{R}^2${ }}\fi}
\def\muO{\relax\ifmmode{\mu_{0}^{2}}\else{$\mu_{0}^{2}${ }}\fi}
\DeclareRobustCommand{\flqq}{\textormath{\guillemotleft}{\mbox{\guillemotleft}}}
\DeclareRobustCommand{\frqq}{\textormath{\guillemotright}{\mbox{\guillemotright}}}%
\def\<{\flqq}\def\>{\frqq}
\newcommand{\myMath}[1]{$#1$}
\renewcommand\thefootnote{\fnsymbol{footnote}}                    
\begin{document}
\thispagestyle{empty}
\vspace*{1mm}

\begin{center}
\textbf{\Large Формфактор пиона в КХД: Как считать?}\\[0.5cm]
 А.~П.~Бакулев$^{\dag}$\\[0.3cm]

\textit{Лаборатория теоретической физики им.~Н.~Н.~Боголюбова, ОИЯИ,\\
    141980 г.~Дубна, Россия}\\[0.9cm]
\end{center}\vspace*{-5mm}

 Обсуждается расчет формфактора пиона,
 $F_\pi(Q^2)$, в КХД.
 Мы кратко обсуждаем основные моменты ПС КХД с нелокальными вакуумными конденсатами
 (НВК) и
 показываем его результаты для  электромагнитного формфактора пиона
 в сравнении с предсказаниями пертурбативной и решеточной КХД.
 В заключение мы рассматриваем подход локальной дуальности (ЛД)
 для формфактора пиона в КХД и
 показываем,
 что при $Q^2\gtrsim 2$~ГэВ$^2$
 основной параметр подхода,
 а именно, порог континуума
 $s_0^{\text{ЛД}}(Q^2)$ должен расти с ростом $Q^2$,
 а не оставаться постоянным.
 \vspace {20mm}
  \renewcommand\thefootnote{\fnsymbol{footnote}}
   \setcounter{footnote}{1}

 \vspace*{1mm}

\begin{center}
\textbf{\Large Pion Form Factor in QCD: How to Calculate?}\\[0.5cm]
Alexander~P.~Bakulev\footnote{E-mail: bakulev@theor.jinr.ru}\\[0.3cm]

\textit{Joint Institute for Nuclear Research,
        Bogoliubov  Lab. of Theoretical Physics,\\
        141980, Moscow Region, Dubna, Russia}\\[0.9cm]
\end{center}

 We discuss the pion form factor calculation in QCD.
 We shortly consider the main points
 of the nonlocal condensate QCD sum rule approach
 and show its results for the pion form factor, $F_\pi(Q^2)$.
 These results are compared with
 predictions of the perturbative and lattice QCD.
 Then we consider the Local Duality (LD) approach
 for the pion FF in QCD and
 show that for $Q^2\gtrsim 2$~GeV$^2$
 the main parameter of the approach,
 namely, $s_0^{\text{LD}}(Q^2)$ should grow
 with $Q^2$ rather than be a constant.
 \vspace*{2mm}

\cleardoublepage
\section{Введение}
 \label{sec:Intro}
 Вычисление асимптотики пионного ФФ \myMath{F_{\pi}(Q^2)}
при \myMath{Q^2\gg 1}~ГэВ$^2$ в пертурбативной КХД~\cite{CZS77-rus,ER80-rus,Rad90}
с самого начала рассматривалось как один из важных успехов КХД
в описании эксклюзивных адронных процессов.
 Однако при использовании пертурбативного подхода КХД-факторизации
для ФФ пиона всегда оставался открытым вопрос о том,
какую часть полного ответа описывает факторизуемый вклад.
Пертурбативный вклад в ФФ пиона
(его часто называют \<жестким\>)
преобладает при асимптотически больших значениях
передач импульса, \myMath{Q^2\gtrsim20-100~\gev{2}}.
Многие авторы считают
(см., например, \cite{Rad84,IL84,JK93,SSK00,BPSS04},
 а также ссылки в этих работах),
что один только жесткий (факторизуемый) вклад в электромагнитный ФФ пиона
слишком мал для объяснения существующих экспериментальных данных
в области умеренных передач \myMath{Q^2=3-10~\gev{2}}
\cite{FFPI78,JLab08II} ---
здесь доминирует так называемый \<мягкий\> вклад.
В области же \myMath{Q^2\lesssim1~\gev{2}} коллинеарное приближение
пертурбативной КХД перестает быть адекватным и
факторизуемая часть ФФ пиона растет как \myMath{1/Q^2},
что никак не согласуется со свойством полного ФФ
\myMath{F_{\pi}(0)=1},
которое диктуется тождеством Уорда.

В нашей работе~\cite{BPSS04} мы показали,
что достаточно разумное описание
может быть получено,
если сшить мягкий вклад,
получаемый в подходе Локальной Дуальности (ЛД)~\cite{NR82,Rad95},
с жестким с использованием самой простой функции включения
пертурбативного вклада.
Слабым местом этого подхода
является получаемая по наследству от ЛД
неопределенность выбора эффективного порога континуума \myMath{s_0^\text{LD}(Q^2)}
при умеренных, но не малых, значениях \myMath{Q^2\geq1~\gev{2}}.
Этот вопрос мы обсудим в разделе~\ref{sec:LD-approach}.

 В этой лекции мы рассматриваем вопрос об учете нефакторизуемых
вкладов в ФФ пиона с помощью нелокальных вакуумных конденсатов (НВК)
в методе правил сумм (ПС) КХД.

Отметим, что ПС с локальными конденсатами
для ФФ пиона обладают плохой стабильностью
и устойчивы лишь при передачах импульса
\myMath{1~\text{ГэВ}^2<Q^2<3~\text{ГэВ}^2}.
Причиной является разное \myMath{Q^2}-поведение
пертурбативного вклада, падающего с ростом \myMath{Q^2},
и непертурбативных вкладов,
которые либо постоянны, либо линейно растут с \myMath{Q^2}~\cite{NR82,IS82}.
Чтобы получить правильные зависимости от \myMath{Q^2},
необходимо вычислять вклады от операторов высшей размерности
типа
\myMath{\langle \bar q(0)D^2q(0)\rangle}, \myMath{\langle \bar q(0)(D^2)^2q(0)\rangle} и т.~п..
Они получаются с помощью тейлоровского разложения изначально нелокальных конденсатов
(напр. \myMath{\langle \bar q(0)q(z)\rangle}).
Полный конденсатный вклад для ФФ убывает с ростом \myMath{Q^2},
в то время как каждый по отдельности
вклад стандартного операторного разложения
имеет структуру \myMath{Q^{2n}}.
Это означает, что для получение осмысленного результата
необходимо отсуммировать весь ряд Тейлора,
что выполнить точно не представляется возможным.
Вариантом такого суммирования и является
метод НВК\footnote{
В работе~\cite{SheSim01} рассматривается операторное разложение
для поляризационного оператора в 3+1-мерной КХД
в подходе фонового поля и делается вывод
об асимптотическом характере получаемого при этом ряда
\myMath{\sum_{n}\lambda_{2n}(4\pi\sigma/Q^2)^n},
где \myMath{\sigma} --- натяжение струны в потенциале конфайнмента,
поскольку коэффициенты \myMath{\lambda_{2n}} растут с \myMath{n} факториально.
Этот вывод показывает, что стремление учесть б\'{о}льшее число
вкладов в операторном разложении не всегда оправдано
--- в асимптотических рядах часто лучше вовремя остановиться.
Реально так и поступают в методе правил сумм КХД:
обычно учитываются лишь вклады глюонного,
\myMath{\va{GG}/Q^4}, и кваркового,
\myMath{\va{\bar{q}q}^2/Q^6}, конденсатов.},
в котором нелокальные вакуумные объекты параметризуются минимальным образом ---
с использованием единственного параметра
\myMath{\lambda_q^2}.

Подход к расчету пионного ФФ методом ПС КХД~\cite{NR82,IS82}
основан на анализе трехточечного \myMath{AAV}-коррелятора
(\myMath{A} --- для аксиального,
 \myMath{V} --- для векторного токов)
с помощью операторного разложения и двойного дисперсионного представления.
В следующем разделе мы разберем этот подход на примере
стандартных ПС КХД и разберем его достоинства и недостатки.

Затем, в разделе~\ref{sec:NLC},
мы систематизируем используемые нами гауссовы модели НВК,
минимальную \cite{MR92,BM98,BMS01},
и улучшенную \cite{BMS01,BP06rus},
и объясним,
чем они отличаются друг от друга:
оказывается, что улучшенная модель
по другому параметризует кварк-глюон-антикварковый НВК,
что позволяет удовлетворить основному КХД-уравнению движения для
векторного билокального НВК и минимизировать нарушение
непоперечности коррелятора векторных токов,
индуцируемое введением нелокальностей вакуумных конденсатов.

В разделе \ref{sec:SR.piFF} мы обсуждаем анализ
полученного ПС КХД с НВК для пионного ФФ,
проводим их сравнение с результатами,
полученными в других теоретических подходах,
в решеточных моделированиях и с экспериментальными данными.

Раздел \ref{sec:LD-approach} посвящен подходу локальной дуальности (ЛД).
Этот подход важен для нас по двум причинам:\\
1) по историческим --- мы пользовались им в работе~\cite{BPSS04}
 для получения предсказаний для полного ФФ пиона методом сшивания
 \<мягкого\> вклада, моделировавшегося в подходе ЛД,
 и \<жесткого\> двухпетлевого вклада;\\
2) по возможности учесть \myMath{O(\alpha_s^2)}-поправку к ФФ пиона
 без счета соответствующей трехпетлевой спектральной плотности.\\
Мы обсуждаем в этом разделе определение эффективного порога
континуума \myMath{s_0^\text{LD}(Q^2)} --- ключевого параметра ЛД
--- при промежуточных значениях \myMath{Q^2} и показываем,
что ранее использовавшееся постоянное значение
\myMath{s_0^\text{LD}(Q^2)\simeq0.6~\gev{2}}
сильно недооценено:
на самом деле, эффективный порог ЛД должен монотонно
расти с ростом \myMath{Q^2} для того, чтобы воспроизводить
результаты борелевских ПС.

В Заключении мы суммируем наши выводы о ФФ пиона.

\section{Стандартные правила сумм КХД для ФФ пиона}
 \label{sec:QCDSR.FF}

 Правила сумм КХД для ФФ пиона строятся
на основе анализа трехточечного \myMath{AAV}-коррелятора
\begin{equation}
 T(p_1^2,p_2^2,q^2)
  =
  \frac{n^{\alpha}n_{\mu}n^{\beta}}{(np_1)^3}
  \int\!\!\!\!\int\!\!d^4x\,d^4y\,e^{i(qx-p_2y)}
  \langle 0|T\!\left[J^{+}_{5\alpha}(y) J^{\mu}(x) J_{5\beta}(0)
               \right]\!|0\rangle,
\label{eq:Corr.JJJ}
\end{equation}
где \myMath{q} --- импульс виртуального фотона (\myMath{q^2=-Q^2}),
а \myMath{p_1} и \myMath{p_2} импульсы налетающего и вылетающего пионов.
В этом корреляторе
\myMath{J^{\mu}(x) = e_u\,\overline{u}(x)\gamma^\mu u(x)
            + e_d\,\overline{d}(x)\gamma^\mu d(x)}
это электромагнитный ток легких кварков,
\myMath{e_u=2/3} и \myMath{e_d=-1/3} --- электрические заряды
\myMath{u}- и \myMath{d}-кварков,
\myMath{J_{5\beta}(x) = \overline{u}(x)\gamma_5\gamma_\beta d(x)}
и
\myMath{J^+_{5\alpha}(x) = \overline{d}(x)\gamma_5\gamma_\alpha u(x)}
--- аксиально-векторные пионные токи,
которые имеют ненулевые проекции на пионное состояние
\myMath{|\pi(P)\rangle}:
\begin{eqnarray}
 \label{eq:Pi.Ax.Curr}
  \langle0|j_5^\alpha|\pi(P)\rangle
   = i\,f_\pi\,P^\alpha\,.
\end{eqnarray}

\begin{figure}[t]
 \centerline{\includegraphics[width=0.8\textwidth]{
  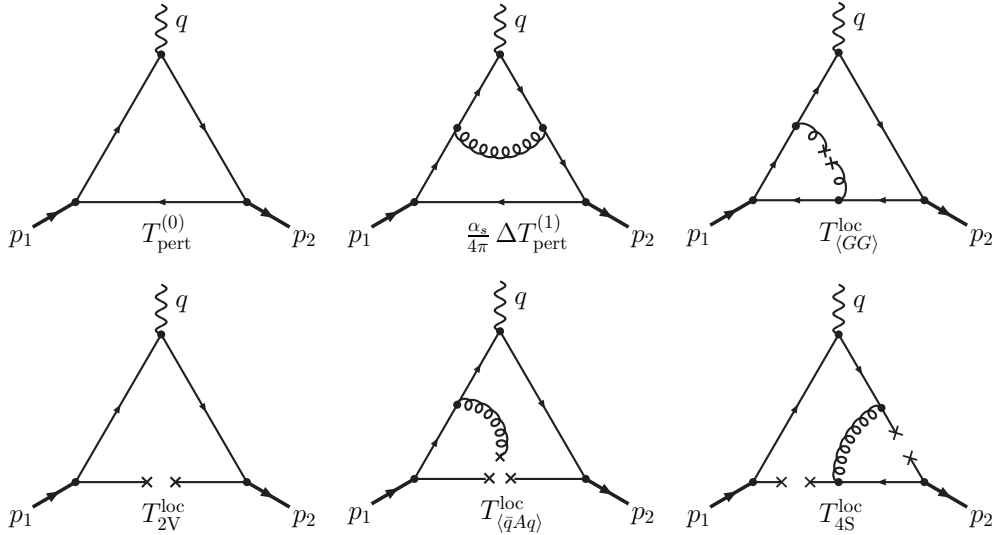}}
  \caption{Диаграммы стандартного операторного разложения трехточечного
    коррелятора \myMath{T(p_1^2,p_2^2,q^2)} с обычными вакуумными конденсатами
    кварковых и глюонных полей (обозначаемых на диаграммах линиями
    с крестиками \myMath{\bm{\times}} на концах).\label{fig:ope-loc}}
\end{figure}

Для этого коррелятора
при достаточно больших пространственно-подобных импульсах
\myMath{p_1} и \myMath{p_2} (\myMath{-p_1^2\gg1~\gev{2}}, \myMath{-p_2^2\gg1~\gev{2}})
строится операторное разложение
(\myMath{c_1} и \myMath{c_2} --- численные константы)
\begin{eqnarray}
 T(p_1^2,p_2^2,q^2)
  = T^\text{pert}(p_1^2,p_2^2,q^2)
  + c_1\frac{\langle GG\rangle}{(p^2)^3}
  + c_2\frac{\alpha_s\langle\bar qq\rangle^2}{(p^2)^4}
  + \ldots\,,
 \label{eq:ope}
\end{eqnarray}
проиллюстрированное на Рис.~\ref{fig:ope-loc},
а также дисперсионное представление
\begin{eqnarray}
 T(p_1^2,p_2^2,q^2)
  = \frac1{\pi^2}
     \int\limits_{0}^{s_0}\!\!\int\limits_{0}^{s_0}\!ds_1\,ds_2\,
       \frac{\rho(s_1,s_2,q^2)}{(s_1-p_1^2)(s_2-p_2^2)}
  + \text{``вычитания''}.
  \label{eq:dr}
\end{eqnarray}
Подобное дисперсионное представление
справедливо и для чисто пертурбативной части коррелятора
\myMath{T^\text{pert}(p_1^2,p_2^2,q^2)}.
Соответствующая трехточечная спектральная плотность
представляется в виде
в \myMath{O(\alpha_s)}-порядке
\begin{eqnarray}
 \label{eq:SpDen.pert}
  \rho^{(1)}_3(s_1, s_2, Q^2)
  \!&\!=\!&\! \left[\rho_3^{(0)}(s_1, s_2, Q^2)
        + \frac{\alpha_s(Q^2)}{4\pi}\,
           \Delta\rho_3^{(1)}(s_1, s_2, Q^2)
    \right]\,.
\end{eqnarray}
Напомним, что в ведущем порядке
спектральная плотность
была рассчитана в работах~\cite{IS82,NR82}
\begin{eqnarray}
 \label{eq:RoSq}
  \rho_3^{(0)}(s_1, s_2, t)
    \!&\!=\!&\!
       \frac{3}{4\pi^2}
       \left[t^2 \frac{d^2}{dt^2}
             + \frac{t^3}{3} \frac{d^3}{dt^3}
       \right]
       \frac{1}{\sqrt{\left(s_1 + s_2 + t\right)^2 - 4\,s_1s_2}}\,,
\end{eqnarray}
в то время как явное выражение спектральной плотности
в следующем за ведущим порядке теории возмущений
\myMath{\Delta\rho_3^{(1)} (s_1, s_2, Q^2)}
было рассчитано совсем недавно~\cite{BO04}.\footnote{
Это выражение достаточно громоздкое, поэтому мы не приводим его здесь,
отсылая заинтересованного читателя к первоисточнику.}
Разница между полной спектральной плотностью \myMath{\rho(s_1,s_2,q^2)}
и ее пертурбативным аналогом
\myMath{\rho^{(1)}(s_1,s_2,q^2)}
компенсируется в операторном разложении (\ref{eq:ope})
непертурбативными вкладами кваркового и глюонного конденсатов,
\myMath{\langle\bar qq\rangle} и \myMath{\langle GG\rangle}.

Вклад высших резонансов, \myMath{T_\text{HR}},
моделируются с использованием той же пертурбативной спектральной плотности
в виде континуума,
\begin{equation}
 \label{eq:HR}
  \rho_\text{HR}(s_1, s_2)
  =  \left[1-\theta(s_1<s_0)\theta(s_2<s_0)\right]\,
      \rho_3(s_1, s_2, Q^2)\,,
\end{equation}
где \myMath{s_0} является эффективным порогом включения высших резонансов.
Обычно \myMath{s_0} соответствует средней точке между массами нижайшего состояния
и первого возбуждения.
Поскольку квадрат массы \myMath{a_1}-мезона \myMath{\simeq1.6~\gev{2}},
ожидаемое значение \myMath{s_0} в случае пиона около \myMath{0.8~\gev{2}}.

В результате, после применения преобразования Бореля,
которое убивает все вычитания в дисперсионном интегралах (\ref{eq:dr})
и подавляет вклад высших резонансов в спектральном интеграле,
мы получаем следующее ПС:
\begin{eqnarray}
 f_\pi^2\,F_\pi(Q^2)
  = \frac1{\pi^2}
     \int\limits_{0}^{s_0}\!\!\int\limits_{0}^{s_0}\!ds_1\,ds_2\,
      \rho^\text{pert}(s_1,s_2,Q^2)e^{-\left(s_1+s_2\right)/M^2}
  + a\,\frac{\alpha_s\langle GG\rangle}{12\pi M^2}
  + b\,\frac{16\pi \alpha_s\langle\bar qq\rangle^2}{81M^4}~~~
 \label{eq:FF.SR.Gen}
\end{eqnarray}
с коэффициентами \myMath{a} и \myMath{b},
определяемыми операторным разложением.
Это ПС анализируется обычным образом:
для каждого заданного значения \myMath{Q^2}
ищется значение порога \myMath{s_0},
при котором значение \myMath{f_\pi^2 F_\pi(Q^2)},
определяемое из ПС (\ref{eq:FF.SR.Gen}),
наименее сильно зависит от параметра Бореля \myMath{M^2}
в определенном интервале значений \myMath{M^2\in[M_{-}^2,M_{+}^2]},
 называемом в подходе ПС окном доверия.
Для пионного ФФ это окно таково:
\myMath{M^2\in[1~\text{ГэВ}^2,2~\text{ГэВ}^2]}.

Существует связь между значением \myMath{f_\pi^2}
и \myMath{F_\pi(Q^2)}
в окне доверия и
параметром \myMath{s_0}.
С достаточной точностью она описывается соотношением локальной дуальности,
следующим из правила сумм (\ref{eq:FF.SR.Gen})
в формальном пределе \myMath{M^2\to \infty}
\begin{eqnarray}
 F_\pi^{\text{LD;}(0)}(Q^2)
  =\frac1{\pi^2f_{\pi}^2}
     \int\limits_{0}^{s_0}\!\!\int\limits_{0}^{s_0}\!ds_1\,ds_2\,
      \rho_3^{(0)}(s_1,s_2,Q^2)
  =
   \frac{s_0}{4\pi^2f_{\pi}^2}
    \left[1-\frac{1+6s_0/Q^2}{(1+4s_0/Q^2)^{3/2}}\right].
 \label{eq:F_pi.LD}
\end{eqnarray}
На первый взгляд конденсаты отсутствуют в этом соотношении.
Однако значение интервала дуальности \myMath{s_0},
извлекаемое в описанной выше процедуре подгонки,
неявно зависит от их величины,
или, точнее, от соотношения конденсатного и пертурбативного вкладов.

Форма операторного разложения по \myMath{1/p^2} трехточечного коррелятора
\myMath{T(p_1^2,p_2^2,q^2)}
зависит от взаимоотношения масштабов \myMath{Q^2} и \myMath{|p^2|}
(под \myMath{p^2} мы подразумеваем оба масштаба:
\myMath{|p|^2\sim|p_1^2|\sim|p_2^2|}).
Простейшей является симметричная ситуация
\myMath{|p^2|\sim Q^2} (промежуточные значения \myMath{Q^2}),
изученная в~\cite{NR82,IS82}.
В этом случае
\myMath{a=1}, \myMath{b=\left(13+2Q^2/M^2\right)}.
Поучительно сравнить эти значения коэффициентов операторного разложения
с коэффициентами,
полученными в ПС КХД для пионной константы распада~\cite{SVZ}
(следует отметить, что значение параметра Бореля в двухточечных ПС,
 \myMath{m^2},
 вдвое меньше по сравнению с трехточечными ПС,
 \myMath{m^2 = M^2/2}):
\begin{eqnarray}
 f_\pi^2
  = \frac{m^2}{4\pi^2}(1-e^{-s_0/m^2})
  + \frac{\alpha_s\langle GG\rangle}{12\pi m^2}
  + 11\frac{16\pi\alpha_s\langle\bar qq\rangle^2}{81m^4}.
 \label{eq:fpi.SR}
\end{eqnarray}
Видно, что ПС для \myMath{f_\pi^2 F_\pi(Q^2)} отличается от ПС
для \myMath{f_{\pi}^2}
удвоенным \myMath{\alpha_s\langle GG\rangle}-вкладом и
примерно учетверённым \myMath{\alpha_s\langle\bar qq\rangle^2}-вкладом
(точнее, этот вклад больше в 3.4 раза).
По этой причине значение \myMath{s_0\approx 4\pi^2 f_{\pi}^2},
диктуемое ПС для \myMath{f_{\pi}^2},
воспроизводится в ФФ ПС
только в области умеренных \myMath{Q^2},
где \myMath{F_\pi(Q^2)} изменяется в пределах \myMath{0.5} и \myMath{0.3},
т.~е. для \myMath{Q^2=0.5-1~\gev{2}},
когда отношение конденсатного и пертурбативного вкладов
в ФФ ПС (\ref{eq:FF.SR.Gen})
очень близко к соответствующему отношению в ПС (\ref{eq:fpi.SR}).

  Это предположение подкрепляется явной обработкой ФФ ПС:
\myMath{s_0=0.7~\gev{2}} для \myMath{Q^2=0.5~\gev{2}}
и \myMath{s_0=0.9~\gev{2}} для \myMath{Q^2=1~\gev{2}}.
Для б\'{о}льших \myMath{Q^2} обработка ПС дает б\'{о}льшие значения порога:
\myMath{s_0=1.0~\gev{2}} для \myMath{Q^2=1.5~\gev{2}},
а для \myMath{Q^2\lesssim0.5~\gev{2}}
мы должны были бы получить значения порога,
меньшие \myMath{0.7~\gev{2}}.
Однако в этой области малых \myMath{Q^2}
необходимо использовать модифицированное операторное разложение\cite{NR84-rus},
включающее дополнительные вклады,
исчезающие при больших значениях \myMath{Q^2}.
В частности, при \myMath{Q^2=0}
конденсатные вклады в ФФ ПС совпадают с аналогичными вкладами
в ПС для \myMath{f_{\pi}^2},
что является следствием тождества Уорда \myMath{F_{\pi}(0)=1},
а затем уменьшаются к значениям,
полученным в симметричным кинематике:
так, например, \myMath{a=(1+m_{\rho}^2/[m_{\rho}^2+Q^2])}.
Следовательно, в области малых \myMath{Q^2}
конденсатные вклады в ФФ ПС согласованы с пертурбативными
в той же степени,
что и в ПС для \myMath{f_{\pi}^2} (\ref{eq:fpi.SR}),
и в результате порог \myMath{s_0} остается близким
исходному значению,
диктуемому тождеством Уорда при  value \myMath{Q^2=0},
а именно,
к \myMath{s_0 \approx 4\pi^2 f_{\pi}^2\approx 0.7~\gev{2}}.

В работе~\cite{NR82} было показано,
что подход ЛД (\ref{eq:F_pi.LD}) с постоянным порогом
\myMath{s_0^\text{LD}(Q^2)=4\pi^2 f_{\pi}^2},
дает вклад достаточно большой для описания экспериментальных данных
вплоть до значений \myMath{Q^2 \sim 2-3~\gev{2}}.
Если же брать растущие с \myMath{Q^2} пороги \myMath{s_0^\text{LD}(Q^2)},
можно получить и б\'{о}льшие значения ФФ.
Мы вернемся к этому вопросу в разделе~\ref{sec:LD-approach}.

Возвращаясь к ФФ ПС (\ref{eq:FF.SR.Gen}),
полученным в~\cite{NR82,IS82},
заметим, что при больших значениях \myMath{Q^2>3~\gev{2}}
они становятся разбалансированными.
Действительно,
пертурбативный вклад спадает как \myMath{s_0/Q^4} или \myMath{M^2/Q^4},
в то время как вклад глюонного конденсата остается неизменным,
а вклад кваркового --- растет линейно по \myMath{Q^2}.
Поэтому для получения физически значимых оценок пионного ФФ
в области промежуточных значений \myMath{Q^2=3-10~\gev{2}}
необходимо серьезно улучшить ПС.

\section{Нелокальные вакуумные конденсаты: гауссовы модели вакуума КХД}
 \label{sec:NLC}

Тот факт, что вклады конденсатов в ПС для ФФ пиона
остаются постоянными или даже растут при увеличении \myMath{Q^2},
достаточно удивителен,
поскольку обычно диаграммы Фейнмана генерируют
\emph{убывающие} с ростом \myMath{Q^2} вклады,
как это происходит с пертурбативным вкладом.
Однако конденсатные диаграммы отличаются от обычных диаграмм Фейнмана
в теории возмущений КХД:
они получаются при замене некоторых пропагаторных линий
постоянными множителями, отвечающими конденсатам:
на рис.\,\ref{fig:ope-loc} такие линии оканчиваются крестиками.

Так, например, кварковый пропагатор
\myMath{\langle T(q(z)\bar q(0))\rangle}
заменяется кварковым конденсатом
\myMath{\langle \bar q(0)q(0)\rangle}
(на рис.\,\ref{fig:ope-loc} такому конденсату отвечает линия
\rule[2.5pt]{15pt}{1pt}\hspace*{-5pt}\myMath{\bm{\times}}
~~\myMath{\bm{\times}}\hspace*{-5pt}\rule[2.5pt]{15pt}{1pt}).
В результате вместо получения зависящего от \myMath{Q^2} результата
получается постоянный вклад.
Зависимость же от \myMath{Q^2} появляется тогда,
когда учитываются вклады операторов старшей размерности типа
\myMath{\langle \bar q(0)D^2q(0)\rangle},
\myMath{\langle \bar q(0)(D^2)^2q(0)\rangle}
и т.~п.,
генерируемые разложением в ряд Тейлора
исходного нелокального вакуумного конденсата (НВК)
\myMath{\langle \bar q(0)q(z)\rangle},
который является непертурбативной частью кваркового пропагатора.

Полный вклад конденсата убывает при больших значениях \myMath{Q^2},
что ясно из общих свойств диаграмм Фейнмана.
Однако \myMath{n}-ый член разложения НВК в ряд Тейлора
имеет поведение типа \myMath{(Q^2/M^2)^n},
и чтобы получить осмысленный,
т.~е. убывающий при больших \myMath{Q^2},
результат
необходимо отсуммировать весь ряд.
Вместо этого мы следуем иной стратегии:
мы не разлагаем НВК в ряд Тейлора,
а пользуемся для них простыми гауссовыми моделями,
учитывающими конечную ширину распределения вакуумных кварков по импульсам,
которые приводят к модифицированной диаграммной технике
с новыми линиями и вершинами, отвечающими НВК.

Вакуумное среднее билокального по кварковым полям оператора
в общем виде можно представить
\begin{eqnarray}
 \langle{\bar{q}^a_A(0)q^b_B(x)}\rangle
  = \frac{\delta^{ab}}{N_c}
     \int\limits_0^\infty
      \left\{\frac{\delta_{AB}}{4}\,
              \langle{\bar{q}q}\rangle\,
              f_S(\alpha)
           - \frac{\widehat{x}_{BA}}{4}\,
              iA_0\,
               f_V(\alpha)
      \right\}
      e^{\alpha x^2/4}\,d\alpha\,,~~~
       \label{eq:va1}
\end{eqnarray}
где \myMath{A_0={2\alpha_s\pi\langle{\bar{q}q}\rangle^2}/{81}},
а \myMath{f_S(\alpha)} и \myMath{f_V(\alpha)} --- функции,
параметризующие скалярный и векторный конденсаты, соответственно.
Для этих конденсатов
мы используем минимальную гауссову модель,
предложенную в~\cite{BM98,BMS01}
и отвечающую выбору
\myMath{f_S(\alpha)=\delta(\alpha-\lambda_q^2/2)}
и \myMath{f_V(\alpha)=\delta'(\alpha-\lambda_q^2/2)}:
\begin{subequations}
\label{eq:Gauss.NLC}
\begin{eqnarray}
\label{eq:S.NLC}
 M_{S}(z^2)
 \!&\!\equiv\!&\!
 \langle{\bar{q}(0)q(z)}\rangle
  = \langle{\bar{q}q}\rangle\,
       e^{-|z^2|\lambda_q^2/8}\,,
\\
\label{eq:V.NLC}
 M_{\mu}(z)
 \!&\!\equiv\!&\!
  \langle{\bar{q}(0)\gamma_\mu q(z)}\rangle
  = \frac{i\, z_\mu\,z^2}{4}\,
       A_0\
        e^{-|z^2|\lambda_q^2/8}\,.
\end{eqnarray}
\end{subequations}
Параметр нелокальности
\myMath{\lambda_q^2 = \langle{k^2}\rangle}
характеризует средний квадрат импульса кварков в вакууме КХД.
Его оценки с помощью стандартных правил сумм КХД~\cite{BI82lam-rus,OPiv88-rus}
и на решетке~\cite{DDM99,BM02} дали следующий интервал возможных значений:
\myMath{\lambda_q^2 = 0.45\pm 0.1~\gev{2}}.

Для векторного и аксиально-векторного кварк-глюон-антикварковых конденсатов
мы используем параметризацию,
предложенную в~\cite{MR89-rus,MR92}:
\begin{subequations}
\begin{eqnarray}
 \langle{\bar{q}(0)\gamma_\mu(-g\widehat{A}_\nu(y))q(x)}\rangle
  \!&\!=\!&\!
  (y_\mu x_\nu-g_{\mu\nu}(y\cdot x))\overline{M}_1(x^2,y^2,(y-x)^2)
  ~~~\nonumber\\
  \!&\!+\!&\!
  (y_\mu y_\nu-g_{\mu\nu}y^2)\overline{M}_2(x^2,y^2,(y-x)^2)\,,~~~
\label{eq:qAq.mu.nu}
 \\
 \langle{\bar{q}(0)\gamma_5\gamma_\mu(-g\widehat{A}_\nu(y))q(x)}\rangle
  \!&\!=\!&\!
   i\varepsilon_{\mu\nu yx}\overline{M}_3(x^2,y^2,(y-x)^2)\,,~~~
\label{eq:qAq.5mu.nu}
\end{eqnarray}
где
\begin{eqnarray}
\label{eq:M_i}
 \overline{M}_i(x^2,y^2,z^2)
  \!&\!=\!&\!
       A_i\int\!\!\!\!\int\limits_{\!0}^{\,\infty}\!\!\!\!\int\!\!
        d\alpha \, d\beta \, d\gamma \,
         f_i(\alpha ,\beta ,\gamma )\,
          e^{\left(\alpha x^2+\beta y^2+\gamma z^2\right)/4}\,,\\
 \label{eq:A_i}
  A_{1,2,3}
  \!&\!\equiv\!&\!
    A_0 \times\left\{-\frac32,2,\frac32\right\}\,.
\end{eqnarray}
\end{subequations}
Функции \myMath{f_i(\alpha,\beta,\gamma)}
можно моделировать по разному.
В минимальной модели~\cite{MR92,BM98,BMS01}
они выбираются в виде произведения \myMath{\delta}-функций
от каждой из переменных типа \myMath{\delta(\alpha-\lambda_q^2/2)},
в то время как в улучшенной модели~\cite{BP06rus}
за счет учета уравнения движения КХД для векторного НВК
получается более сложное выражение.

\begin{figure}[b]
 \centerline{\includegraphics[width=1\textwidth]{
  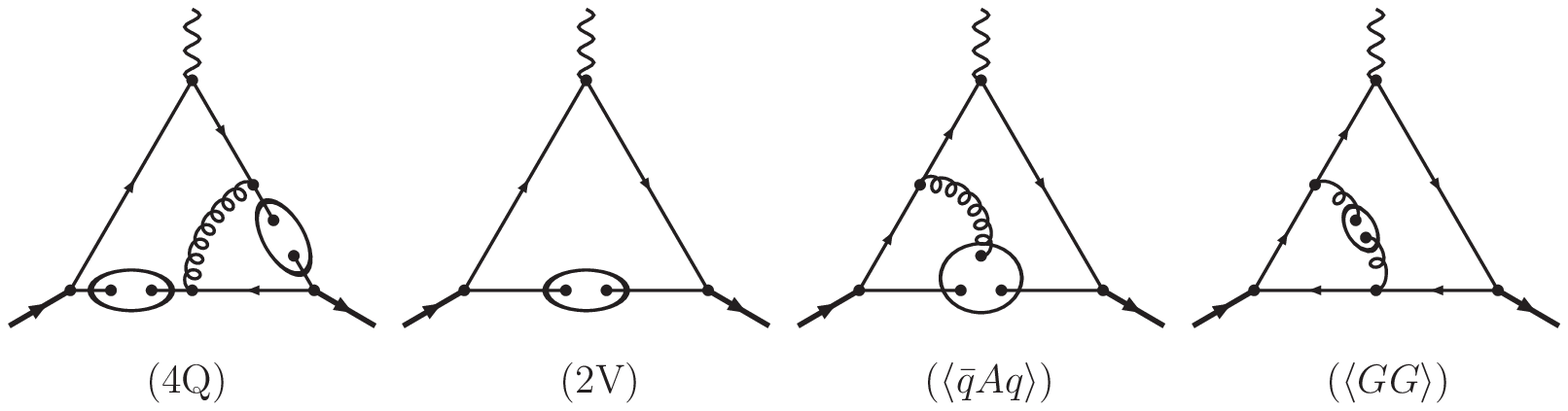}}
  \caption{Непертурбативные вклады~\cite{NR82,IS82,BR91} в правило сумм
   для ФФ пиона, (см.\ (\ref{eq:ffQCDSR})).\label{fig:JJJ.NLC}}
\end{figure}
В результате мы получаем модифицированное операторное разложение
борелевского образа амплитуды (\ref{eq:Corr.JJJ})
\begin{eqnarray}
 \label{eq:Borel.Phi}
  \Phi\left(Q^2,M^2\right)
   \!&\!\equiv\!&\!
    B_{-p_1^2\to M^2,-p_2^2\to M^2}T(p_1^2,p_2^2,-Q^2)
 \nonumber \\
 \label{eq:Borel.Phi.OPE}
   &\!=\!&
    \Phi_\text{pert}(Q^2,M^2)
     + \Phi_{\va{GG}}(Q^2,M^2)
     + \Phi_{\langle\bar{q}q\rangle}(Q^2,M^2)\,,
\end{eqnarray}
где \myMath{\Phi_{\va{GG}}(Q^2,M^2)} обозначает вклад глюонного,
а вклад нелокальных кварковых конденсатов
представляется в виде
суммы четырехкваркового (4Q),
билокального векторного (2V)
и кварк-глюон-антикваркового (\myMath{\bar{q}Aq}) НВК:
\begin{eqnarray}
 \label{eq:Borel.Phi.qq}
  \Phi_{\langle\bar{q}q\rangle}(Q^2,M^2)
    &\!=\!&
     \Phi_\text{4Q}(Q^2,M^2)
    +\Phi_\text{2V}(Q^2,M^2)
    +\Phi_{\va{\bar{q}Aq}}(Q^2,M^2)
\end{eqnarray}
Графическая иллюстрация такого представления приведена
на Рис.\,\ref{fig:JJJ.NLC}:
показаны только типичные диаграммы для каждого подкласса диаграмм,
тогда как полный набор диаграмм включает также зеркально-сопряженные
(для подклассов 4Q и \myMath{\bar{q}Aq}) диаграммы,
а также диаграммы с перестановками вставок глюонных линий
(подкласс G).

Простейшим здесь является вклад векторного конденсата \myMath{M_{\mu}}:
\begin{eqnarray}
 \Phi_\text{2V}(M^2,Q^2)
  = \frac{16A}{M^4} \int_0^1 dx\,
     \bar x\,f_V\left(x\,M^2\right)
      \exp\left(-\frac{xQ^2}{2\bar x M^2}\right)~~~ \nonumber \\
 \label{eq:Phi.V.Gauss}
  \stackrel{\text{Gauss}}{\longrightarrow}
   \frac{8\,A_0}{M^4}\,
    \left(2+\frac{Q^2}{M^2(1-2\,\Delta(M^2)}\right)
     \exp\left[\frac{-\,Q^2\,\Delta(M^2)}
                    {M^2\left(1-2\,\Delta(M^2)\right)}
         \right],
\end{eqnarray}
где \myMath{\Delta(M^2)\equiv\lambda_q^2/(2\,M^2)}.
Как и ожидалось,
он является убывающим для больших \myMath{Q^2},
причем чем больше значение параметра нелокальности вакуума \myMath{\lambda_q^2},
тем быстрее он убывает с ростом \myMath{Q^2}.
Значение \myMath{Q^2_{*}},
при котором начинается убывание,
сильно зависит от значения от значения борелевского параметра \myMath{M^2}.
Принимая значение параметра нелокальности вакуума
\myMath{\lambda_q^2=0.4~\gev{2}},
получаем для \myMath{M^2=1, 1.5, 2~\gev{2}}
следующие значения \myMath{Q^2_{*}=2, 6, 13~\gev{2}},
соответственно.
В локальном же пределе этот вклад равен
\begin{eqnarray}
 \Phi_\text{2V}^\text{loc}(M^2,Q^2)
  = \frac{16\,A_0}{M^8}\,
     \left(2+\frac{Q^2}{M^2}\right)\,.
 \label{eq:Phi.V.loc}
\end{eqnarray}

Скалярный четырехкварковый НВК дает более сложное выражение:
\begin{eqnarray}
 \Phi_\text{4Q}(M^2,Q^2)
  =
   \frac{288 A_0}{M^8}
    \int\limits_{-1}^{1}\!\!\!\!
     \int\limits_{0}^{1}\!\!\!\!
      \int\limits_{0}^{1}\!\!\!\!
       \int\limits_{0}^{1}\!\!\!
        \frac{(a+1)\,\bar{b}\,x\,\bar{y}\,da\,db\,dx\,dy}
             {(a+1)\bar{b}x\bar{y}-\bar{a}(b+1)\bar{x}y}\,
         \theta\left[(a-b)(x-y)\right]
 \nonumber\\
  \times
   \varphi_S\left(\frac{M^2}{\bar{b}},\bar{x}\right)
    \varphi_S\left(\frac{M^2}{a+1},y\right)
     \exp\!\!\left[\frac{-Q^2\,y\left[a\bar{b}x\bar{y}-\bar{a}b\bar{x}y\right]}
                    {M^2\,\bar{y}\left[(a+1)\bar{b}x\bar{y}-\bar{a}(b+1)\bar{x}y\right]}
              \right]
      ,~~~
 \label{eq:Phi.S}
\end{eqnarray}
где мы ввели обезразмеренные функции распределения
\myMath{\varphi_S(M^2,x)=M^2\,f_S(M^2x)},
которые для гауссова приближения (\ref{eq:Gauss.NLC})
сводятся просто к
\myMath{\varphi_S^\text{Gauss}(M^2,x)=\delta(x-\Delta(M^2))}.
Конечно, они снимают два интегрирования и упрощают выражение (\ref{eq:Phi.S}),
но все равно интегралы явно не берутся
и оценивать результат приходится
численным интегрированием.
Отметим, что в локальном пределе этот вклад равен
\begin{eqnarray}
 \Phi_\text{4Q}^\text{loc}(M^2,Q^2)
  = \frac{144\,A_0}{M^8}\,,
 \label{eq:Phi.S.loc}
\end{eqnarray}
и является максимальным среди вкладов кварковых конденсатов
при \myMath{Q^2\approx M^2}.

В локальном пределе кварк-глюон-антикварковый НВК дает вклад
точно такой же,
как и векторный билокальный НВК, см. (\ref{eq:Phi.S.loc}),
что составляет \myMath{1/3} от вклада
\myMath{\Phi_\text{4Q}^\text{loc}(M^2,Q^2)}
при \myMath{Q^2\approx M^2}.
Однако его расчет оказывается самым сложным
и ответ получается достаточно громоздким:
\begin{eqnarray}
 \Phi_{\va{\bar{q}Aq}}(M^2,Q^2)
  =
   \frac{16\,A_0}{M^8}\!
    \int\limits_{0}^{1}\!\!\!\!
     \int\limits_{0}^{1}\!\!\!\!
      \int\limits_{0}^{1}\!\!\!\!
       \int\limits_{0}^{1}\!\! da\,db\,dx\,dy\,
        \exp\left[
         \frac{-Q^2}{M^2}\,
         \frac{\bar{a}\bar{y}+2a\bar{x}-(a+1)\bar{x}\bar{y}}{2(a+1)\bar{x}\bar{y}}
            \right]
 \nonumber\\
  \times
        \theta(y-x)
         \sum\limits_{i=1}^{3}
          a_i\,\left[\varphi_iP_i\left(M^2,a,b,x,y\right)
                   + \bar{\varphi}_iS_i\left(M^2,a,b,x,y\right)
               \right]
     ,~~~~~~~~~~
 \label{eq:Phi.qAq}
\end{eqnarray}
Здесь численные коэффициенты
\myMath{\left\{a_1,a_2,a_3\right\}=\left\{-\frac{3}{4},1,+\frac{3}{4}\right\}}
наследуют иерархическую информацию из коэффициентов
\myMath{A_i}, см. (\ref{eq:A_i}),
а обезразмеренные функции распределения определяются следующим образом:
\begin{subequations}
\begin{eqnarray}
 \label{eq:varphi.i}
  \varphi_i\left(M^2,a,b,x,y\right)
   \!&\!=\!&\!
    \frac{M^6\,\bar{x}}{2\,(1+a)\,\bar{a}}\,
     f_i\left(\frac{M^2\,x}{2},
              \frac{M^2\,\bar{x}\,b}{\bar{a}},
              \frac{M^2\,(y-x)}{a+1}
        \right),~~~\\
 \label{eq:bar-phi.i}
  \bar{\varphi}_i\left(M^2,a,b,x,y\right)
   \!&\!=\!&\!
    \frac{M^6\,\bar{x}}{2\,(1+a)\,\bar{a}}\,
     f_i\left(\frac{M^2\,x}{2},
              \frac{M^2\,(y-x)}{a+1},
              \frac{M^2\,\bar{x}\,b}{\bar{a}}
        \right).~~~
\end{eqnarray}
\end{subequations}
В локальном приближении (\ref{eq:Gauss.NLC})
они сводятся просто к произведению \myMath{\delta}-функций:
\begin{eqnarray}
 \label{eq:varphi.i.loc}
  \varphi_i^\text{loc}\left(M^2,a,b,x,y\right)
   =
  \bar{\varphi}_i^\text{loc}\left(M^2,a,b,x,y\right)
   = \delta(x)\,\delta(y)\,\delta(b)\,.
\end{eqnarray}
Коэффициентные функции \myMath{P_i\left(M^2,a,b,x,y\right)}
и \myMath{S_i\left(M^2,a,b,x,y\right)}
были рассчитаны в нашей работе~\cite{BR91}
и приведены в Приложении~\ref{App:QCDSR.Param},
см. (\ref{eq:qAq.SR.Pi})--(\ref{eq:qAq.SR.Si}).
Эти формулы для коэффициентных функций
\myMath{P_i} и \myMath{S_i}
совместно с конкретной моделью нелокального вакуума КХД,
т.~е. с конкретными модельными функциями
\myMath{\varphi_i(M^2,a,b,x,y)}
позволяют нам получить вклад кварк-глюон-антикваркового НВК
в ПС для ФФ пиона.

\begin{figure}[b!]
 \centerline{\includegraphics[width=0.33\textwidth]{
  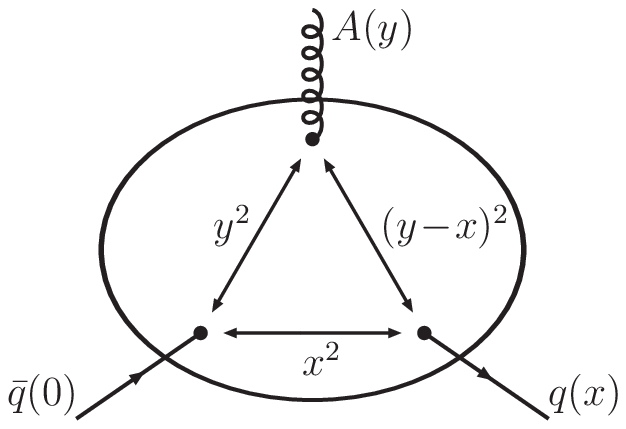}}
   \caption{Иллюстрация координатных зависимостей
   кварк-глюон-анти\-кваркового конденсата.\label{fig:3L}}
\end{figure}
  Отметим, что первоначальный анализ полученных
в нашей работе~\cite{BR91} ПС
был основан на не полностью нелокальной модели вакуума КХД
(\ref{eq:fi.barqAq.BR91})
для кварк-глюон-антикваркового НВК
\myMath{M_i(x^2,y^2,(x-y)^2)},
определяемых по формуле (\ref{eq:M_i})).
В этой модели нелокальность вводилась только для двух
(скажем, \myMath{x^2} и \myMath{(x-y)^2})
из трех имеющихся межпартонных расстояний,
\myMath{x^2}, \myMath{y^2} и \myMath{(x-y)^2} ,
см. Рис.\,\ref{fig:3L}.
Более точно,
использовались следующие параметризации
(\myMath{\Lambda=\lambda_q^2/2}):
\begin{eqnarray}
\label{eq:fi.barqAq.BR91}
 f^\text{BR}_i\left(\alpha,\beta,\gamma\right)
  \!&\!=\!&\!
         \delta\left(\alpha-x_{i1}\Lambda\right)
         \delta\left(\beta -x_{i2}\Lambda\right)
         \delta\left(\gamma-x_{i3}\Lambda\right)\,,
\\\nonumber
 x_{ij}
\!&\!=\!&\!
   \left(
        \begin{array}{ccc}
           0.4 & 0   & 0.4\\
           0   & 1   & 0.4\\
           0   & 0.4 & 0.4
        \end{array}
  \right)\,.
\end{eqnarray}
Нулевые элементы в матрице \myMath{x_{ij}}
говорят об отсутствии нелокальности
либо по кварк-антикварковому расстоянию \myMath{y^2} (\myMath{i=1, j=2}),
либо по антикварк-глюонному расстоянию \myMath{x^2} (\myMath{i=2, 3} и \myMath{j=1}),
см. Рис.\,\ref{fig:3L}.
Таким образом, ПС для ФФ пиона и в такой модели будет частично
неустойчиво, как это происходит в стандартном подходе ПС КХД.

\section{Анализ ПС КХД с НВК для ФФ пиона}
 \label{sec:SR.piFF}
 Для анализа построенных ПС с НВК
\begin{eqnarray}
 \label{eq:ffQCDSR}
  f_{\pi}^2\,F_{\pi}(Q^2)
  &\!=\!&
  \int\limits_{0}^{s_0}\!\!\int\limits_{0}^{s_0}\!ds_1\,ds_2\
           \rho_3(s_1, s_2, Q^2)\,
           e^{-(s_1+s_2)/M^2}
     + \Phi_{\va{GG}}(Q^2,M^2)
     + \Phi_{\langle\bar{q}q\rangle}(Q^2,M^2)~~~~~
\end{eqnarray}
мы используем обе гауссовы модели нелокального вакуума КХД ---
минимальную (\ref{eq:Min.Anz.qGq}) \cite{MR92, BM98,BMS01}
и улучшенную (\ref{eq:Imp.Anz.qGq}) \cite{BP06rus}.
Использование полностью нелокальных моделей вакуума
позволяет нам значительно расширить область применимости ПС КХД
вплоть до передач импульса \myMath{\simeq 10~\gev{2}}.
Кроме того, мы учитываем в пертурбативной спектральной плотности
\myMath{O(\alpha_s)}-поправку \cite{BO04},
что повышает наши предсказания примерно на \myMath{20\%}
по сравнению с оценками работ
\cite{BR91,IS82,NR82},
где была учтена спектральная плотность ведущего порядка.

Имея в виду в дальнейшем использование ДАТВ для оценки вклада
\myMath{O(\alpha_s^2)}-поправки,
см. раздел~\ref{sec:LD-approach},
мы с самого начала применяем однопетлевой аналитический заряд
\myMath{\mathcal A_1^{(1);\text{\tiny glob}}(Q^2)}
с трехфлейворным масштабом \myMath{\Lambda_3=300}~МэВ,
см.~\cite{AB08rus}.

Стратегия численной обработки ПС такая же,
как и в обычном подходе.
Именно, при каждом заданном значении \myMath{Q^2=1-10~\gev{2}}
ПС (\ref{eq:ffQCDSR}) дает нам ФФ пиона
\myMath{F_{\pi}(Q^2,M^2,s_0)}
в виде функции двух вспомогательных параметров ---
борелевского параметра \myMath{M^2}
и эффективного порога континуума \myMath{s_0}.
Параметр \myMath{s_0} определяет границу между пионом
и высшими резонансами в аксиальном канале
(\myMath{A_1}, \myMath{\pi'} и т.~д.)
и поэтому мы предполагаем,
что он не может быть меньше 0.6~ГэВ\myMath{{}^2}.
Конкретное значение \myMath{s_0(Q^2)}
при заданном значении \myMath{Q^2} определяется из требования
минимальной чувствительности функции \myMath{F_{\pi}(M^2,s_0)}
к значению параметра Бореля \myMath{M^2}
внутри окна доверия ПС.
Мы берем эти интервалы доверия и соответствующие им значения пионной константы
\myMath{f_{\pi}}
из двухточечных ПС КХД с соответствующей гауссовой моделью НВК
--- минимальной или улучшенной~\cite{BP06rus}.
Границы интервалов доверия двухточечных ПС с нелокальными конденсатами,
\myMath{M_{\pm}^{2}/2},
внутри которых (\myMath{M^2\in[M^2_{-}/2, M^2_{+}/2]})
предсказаниям ПС можно доверять,
и значения соответствующей им пионной константы распада, \myMath{f_{\pi}},
для двух используемых моделей НВК указаны в Таблице~\ref{eq:CI.and.fpi}.
\begin{table}[t!]
 \caption{Границы интервалов доверия двухточечных ПС
  и соответствующие им значения пионной константы распада
  для двух используемых моделей НВК.
   \label{eq:CI.and.fpi}\vspace*{+1mm}}
  \begin{tabular}{|l|ccc|} \hline
   Модель       & $M^2_{-}$ & $M^2_{+}$   & $f_\pi$   \\ \hline\hline
   Минимальная~\cite{BMS01}
                & 1~ГэВ$^2$ & 1.7~ГэВ$^2$ & 137~МэВ\\
   Улучшенная~\cite{BP06rus}
                & 1~ГэВ$^2$ & 1.9~ГэВ$^2$ & 142~МэВ\\ \hline
  \end{tabular}
\end{table}

Заметим здесь,
что значения параметра Бореля \myMath{M^2} в трехточечных ПС
в первом приближении в два раза больше величины параметра Бореля
в двухточечных ПС:
\myMath{M_\text{3-point}^2=2M_\text{2-point}^2}.
Благодаря положительности спектральной плотности пертурбативного вклада
оказывается,
что чем больше значение \myMath{s_0}
тем больше получаемое значение ФФ пиона.

Используя среднеквадратичное отклонение от среднего в окне доверия,
\myMath{\chi^2(Q^2,s_0)},
см. Приложение~\ref{App:QCDSR.Param}, (\ref{eq:xi}),
мы определяем такой порог континуума
\myMath{s_0^{\text{SR}}(Q^2)},
который минимизирует зависимость правой части (\ref{eq:ffQCDSR})
от параметра Бореля \myMath{M^2\in[M^2_{-}, M^2_{+}]}
при каждом заданном значении \myMath{Q^2}.

\begin{figure}[b!]
 \centerline{\includegraphics[width=0.47\textwidth]{
  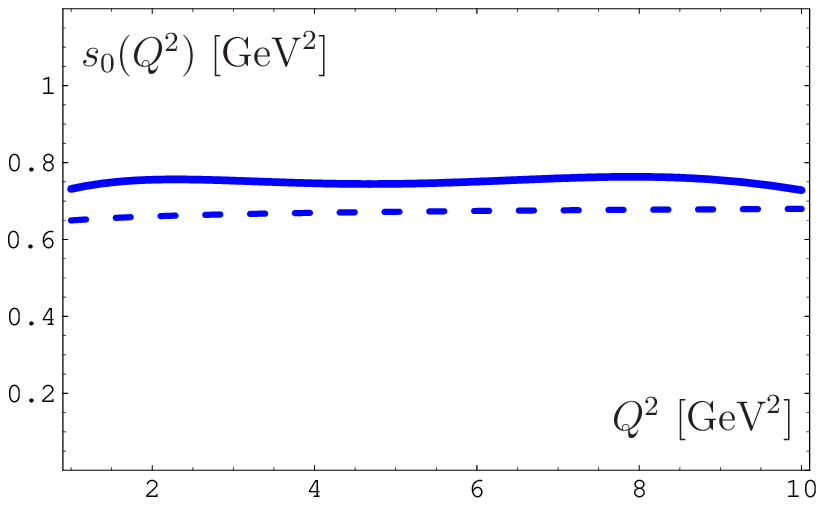}~\includegraphics[width=0.47\textwidth]{
  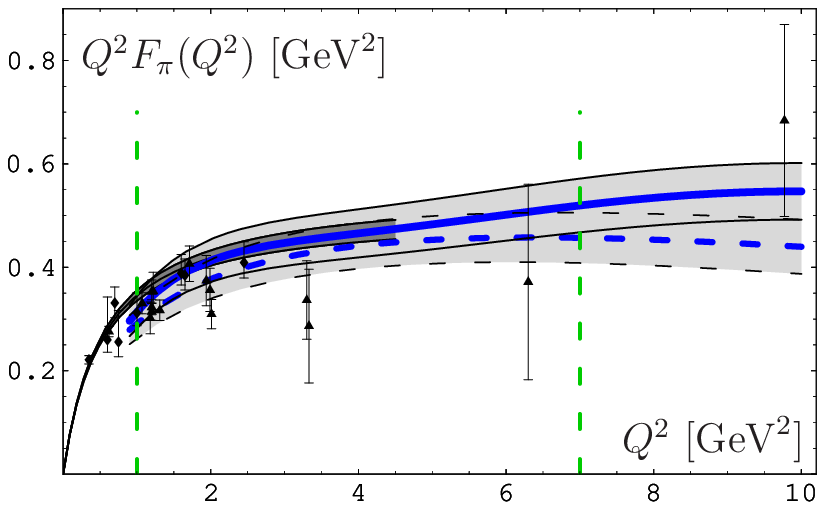}}
   \caption{Слева: Зависимость порога континуума \myMath{s_0(Q^2)~[\gev{2}]}
    для минимальной (штрихованная линия) и
    для улучшенной (сплошная линия) моделей НВК.
    Справа: Нормированный пионный ФФ \myMath{Q^2 F_{\pi}(Q^2)}
    для минимальной (штрихованная линия) и
    для улучшенной (сплошная линия) моделей НВК
    для \myMath{\lambda_q^2=0.4~\gev{2}} в сравнении с экспериментальными
    данными Корнелла \protect\cite{FFPI74,FFPI76,FFPI78} (треугольники)
    и лаб.~им.Джефферсона\cite{JLab08II} (ромбы).\label{fig:main-result}}
\end{figure}

А вот при \myMath{Q^2\gtrsim4~\gev{2}} минимум
во всем интервале допустимых значений \myMath{s_0} отсутствует,
так что невозможно определить порог \myMath{s_0^{\text{SR}}}
таким методом.
Заметим, однако, что  значения \myMath{\min\limits_{s}[\chi^2(Q^2,s)]}
и \myMath{\chi^2(Q^2,s_0=s_0^{\text{LD};(1)}(Q^2)\simeq0.63~\text{ГэВ}^2)}
очень близки:
относительная разность для
\myMath{Q^2=4-10~\gev{2}} имеет порядок
\myMath{10-15}\%\footnote{%
Здесь порог \myMath{s_0^{\text{LD};(1)}(Q^2)} задан стандартным предписанием
приближения локальной дуальности,
см. также обсуждение после  (\ref{eq:LD.s0}).}.
Поэтому в случае минимально гауссовой модели НВК
мы будем использовать \myMath{s_0^{\text{SR}}(Q^2)=s_0^{\text{LD};(1)}(Q^2)}
как порог континуума.
Эти пороги \myMath{s_0^{\text{SR}}(Q^2)} для минимальной (штрихованная линия)
и для улучшенной (сплошная линия) гауссовых моделей НВК показаны
на левой панели Рис.\,\ref{fig:main-result}.

Оценив эффективные пороги континуума,
мы уже можем определить предсказание нашего ПС для пионного ФФ
как среднее значение правой части (\ref{eq:ffQCDSR})
по параметру Бореля \myMath{M^2\in[M^2_{-}, M^2_{+}]}:
\begin{eqnarray}
 \label{eq:FF.pi.SR.result}
 F_{\pi}^\text{SR}(Q^2)
  \!&\!=\!&\! \frac{1}{M_{+}^2-M_{-}^2}
       \int_{M_{-}^2}^{M_{+}^2}
        F(Q^2, M^2, s_0^\text{SR}(Q^2))\,dM^2\,.
\end{eqnarray}
Полученные при этом результаты для обеих исследуемых моделей
нелокального вакуума КХД
(с одинаковым параметром нелокальности \myMath{\lambda_q^2=0.4~\gev{2}})
показаны на Рис.\,\ref{fig:main-result}
в виде двух затененных полос,
ограниченных сплошными и штрихованными линиями,
которые показывают неопределенности предсказаний в,
соответственно, улучшенной и минимальной моделях НВК.

\begin{figure}[b!]
 \centerline{\includegraphics[width=0.47\textwidth]{
  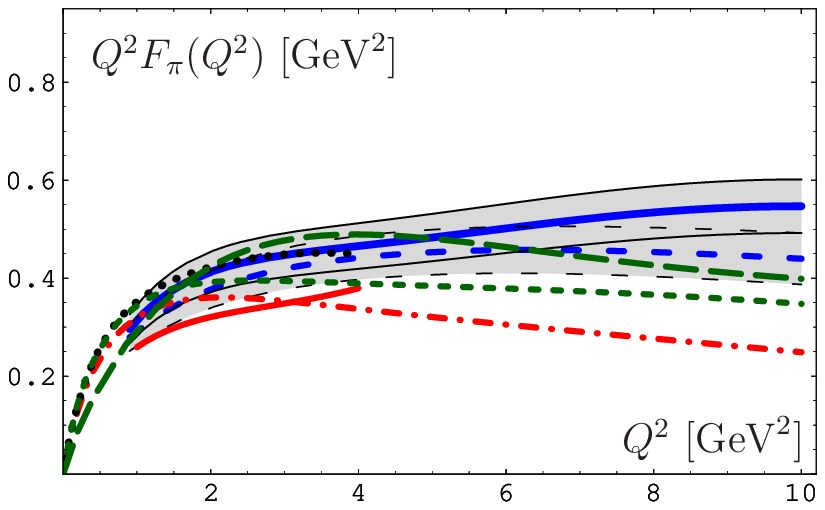}~\includegraphics[width=0.47\textwidth]{
  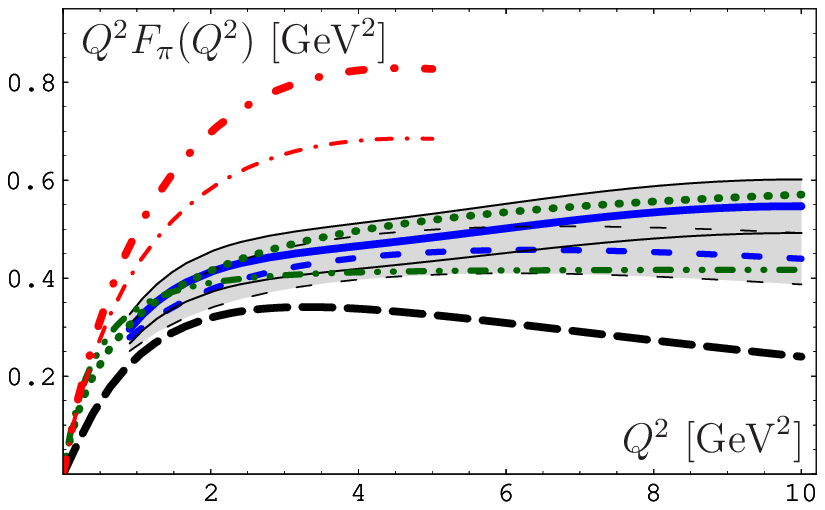}}
  \caption{Сравнение предсказаний для пионного ФФ \myMath{Q^2 F_{\pi}(Q^2)},
   полученных в разных теоретических подходах. Наши результаты показаны
   на обоих панелях теми же затененными полосами, что и
   на Рис.~\ref{fig:main-result}.\label{fig:Q2FFall}}
\end{figure}

Для центральных кривых
(штрихованная отвечает минимальной модели,
 сплошная --- улучшенной)
на этом рисунке,
а также на обеих панелях Рис.\,\ref{fig:Q2FFall},
мы использовали следующие интерполяционные формулы:
\begin{subequations}
\begin{eqnarray}
 \label{eq:FF.pi.SR.Int.Min}
  F_{\pi;\text{min}}^{\text{SR}}(Q^2=x~\text{ГэВ}^2)
   \!&\!=\!&\! e^{-1.402\,x^{0.525}}
        \left(1+0.182\,x+\frac{0.0219\,x^3}{1+x}
        \right),~~~~~\\
 \label{eq:FF.pi.SR.Int.Imp}
  F_{\pi;\text{imp}}^{\text{SR}}(Q^2=x~\text{ГэВ}^2)
   \!&\!=\!&\! e^{-1.171\,x^{0.536}}
        \left(1+0.0306\,x+\frac{0.0194\,x^3}{1+x}\right),~~~~~
\end{eqnarray}
\end{subequations}
справедливые при \myMath{Q^2\in[1,10]~\gev{2}},
т.~е. при \myMath{x\in[1,10]}.
Две вертикальные разрывные линии на Рис.\,\ref{fig:main-result}
обозначают область сильного доверия построенного ПС КХД с НВК:
в этой области предсказания,
полученные на основе двух разных моделей нелокального вакуума КХД
--- минимальной и улучшенной ---
перекрываются.
На границе этой области вблизи \myMath{Q^2\simeq7~\gev{2}}
центральная линии обеих полосок начинают выходить за пределы
полосы предсказаний другой модели.
Однако предсказания сами по себе остаются осмысленными
вплоть до значения \myMath{Q^2\simeq10~\gev{2}}:
при б\'{о}льших значениях \myMath{Q^2}
окно доверия начинает сужаться,
так что ошибки метода нарастают.

 Недавние результаты,
полученные в решеточной КХД \cite{Brommel06},
показаны в виде темной полосы
между двумя жирными сплошными линиями
при малых \myMath{Q^2\lesssim4~\gev{2}}.
Видно достаточно хорошее согласие как с этими псевдоэкспериментальными данными,
так и с реальными экспериментальными данными,
полученными в Корнелле \cite{FFPI74,FFPI76,FFPI78} (треугольники)
и лаб. им.~Джефферсона \cite{JLab08II} (ромбы).

На левой панели Рис.\,\ref{fig:Q2FFall} наши результаты
показаны в сравнении с предсказаниями различных теоретических моделей:
зеленая длинно-штрихованная линия
представляет результаты работы \cite{RA09},
полученные с учетом радиационных \myMath{O(\alpha_s)}-поправок
и вклада твиста 3
в подходе пересуммированной пертурбативной КХД,
в то время как зеленая линия с короткой штриховкой
--- реджевской модели в пределе больших \myMath{N_c} \cite{RAB08}.
Короткая красная жирная сплошная кривая,
оканчивающаяся вблизи точки \myMath{Q^2=4~\gev{2}},
отражает результаты стандартных ПС КХД
с локальными конденсатами~\cite{NR82,IS82},
а красная штрих-пунктирная линия --- недавние оценки,
полученные в подходе ПС с локальной дуальностью \cite{BLM07}.
Отметим здесь, что наши прежние результаты~\cite{BPSS04},
не показанные на этом рисунке,
приблизительно на \myMath{20}\% выше этой штрих-пунктирной линии:
это связано как с учетом \myMath{O(\alpha_s^2)}-поправок (10\%),
так и с неопределенностями процедуры
сшивания \<мягкого\> и \<жесткого\> вкладов,
использовавшейся в \cite{BPSS04}, (10\%)
(см. обсуждение этого вопроса в  разделе\,\ref{sec:LD-approach}
 и на Рис.\,\ref{fig:Q2FFLDvsMOD}).
Наконец, короткий черный пунктир, идущий до точки 4~ГэВ\myMath{{}^2},
показывает результаты модели \cite{MT00},
основанной на уравнении Бете--Солпитера.

На правой панели Рис.\,\ref{fig:Q2FFall} наши результаты
показаны в сравнении
с предсказаниями голографических моделей,
основанных на представлении о дуальности AdS--КХД.
Жирная черная длинно-штрихованная линия показывает результаты,
полученные на основе пионной АР,
извлеченной в AdS--КХД-подходе \cite{AN08},
в то время как зеленая штрих-пунктир-пунктирная линия
--- результаты модели Григоряна--Радюшкина \cite{GR08}.
Пунктирная зеленая линия дает предсказания модели AdS--КХД с мягкой стенкой
 \cite{BT07}.
Наконец, две верхние красные штрих-пунктирные линии
показывают результаты,
полученные в улучшенных моделях AdS--КХД с мягкой
(жирный штрих-пунктир)
и жесткой (тонкий штрих-пунктир) стенками~\cite{KL08},
соответственно.
Видно, что зеленые пунктирная и штрих-пунктир-пунктирная кривые
хорошо согласуются с нашими результатами,
причем предпочтение мы отдаем результатам работы \cite{GR08},
поскольку они оставляют место для радиационных поправок.

\begin{figure}[b!]
 \centerline{\includegraphics[width=0.49\textwidth]{
  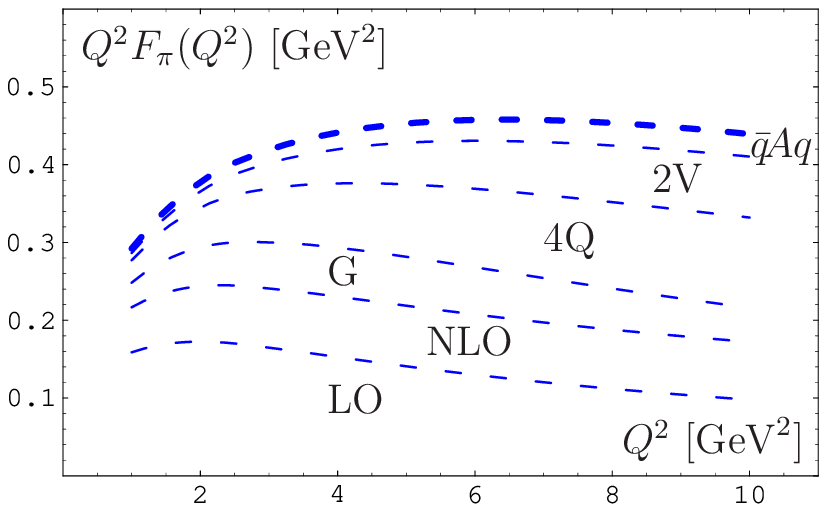}~\includegraphics[width=0.49\textwidth]{
  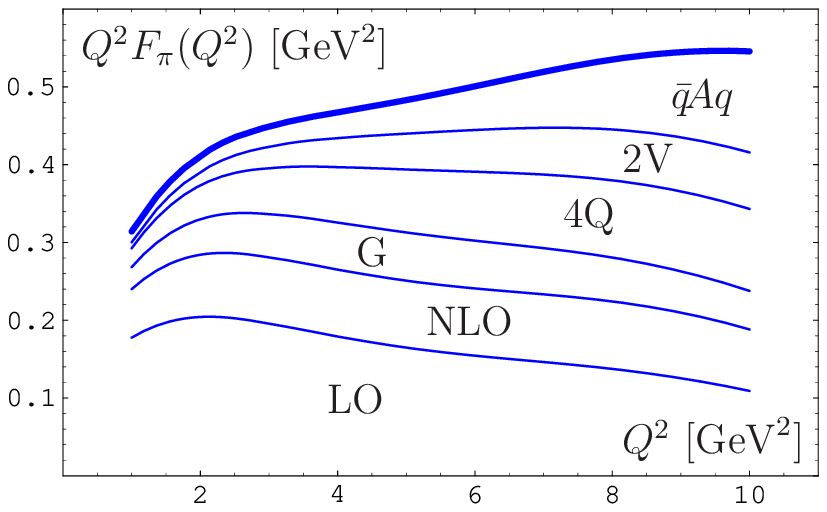}}
   \caption{Сравнение различных вкладов в пионный ФФ в минимальной (слева)
     и улучшенной гауссовых моделях вакуума КХД. Внизу показаны пертурбативные
     \myMath{O\left(1\right)}-вклады (метка LO под кривыми), над ними идут
     кривые, отвечающие сумме \myMath{O\left(1\right)}- и
     \myMath{O\left(\alpha_s\right)}-вкладов (метка NLO), а далее по тому же
     принципу показаны добавки непертурбативных вкладов от:
     четырехкваркового скалярного конденсата (4Q),
     билокального векторного конденсата (2V),
     кварк-глюон-антикваркового конденсата (\myMath{\bar{q}Aq})
     и глюонного конденсата (G).\label{fig:Q2FFapart}}
\end{figure}
На Рис.\,\ref{fig:Q2FFapart}
мы иллюстрируем различия предсказаний для пионного ФФ
в минимальной (слева, штрихованные линии)
и улучшенной (справа, сплошные линии)
гауссовых моделях вакуума КХД.
Различие между этими моделями связано,
как мы уже говорили, с вкладом кварк-глюон-антикваркового конденсата
\myMath{\Phi_{\bar qAq}(Q^2,M^2)},
что хорошо видно по верхним кривым на обеих панелях рисунка.
Кроме того,
это различие влияет на значение порога континуума
\myMath{s_0(Q^2)},
что также приводит к различию пертурбативных вкладов
(LO и NLO).
На рисунке показано каким образом различные вклады суммируются
в конечный ответ (верхние жирные линии на каждой панели):
каждая из кривых на обеих панелях является суммой
(снизу вверх) всех предшествующих вкладов.

Отметим, что все расчеты были сделаны нами
для значения параметра нелокальности,
равного \myMath{\lambda_q^2=0.4~\gev{2}},
которое оказывается выделенным анализом \cite{BMS02,BMS03}
данных CLEO по ФФ фотон-пионного перехода.
Использование б\'{о}льших значений этого параметра нелокальности
привело бы к уменьшению наших предсказаний для электромагнитного ФФ пиона
за счет более сильных эффектов нелокальности
(заметим, что вклады в ПС для пионного ФФ от НВК
 пропорциональны \myMath{\exp[-Q^2\lambda_q^2/M^4]},
 см.~(\ref{eq:Phi.V.Gauss})).
Очевидно, что меньшие значения \myMath{\lambda_q^2}
наоборот повысят значения ФФ пиона в нашем подходе.
Таким образом, наши предсказания на Рис.\,\ref{fig:main-result},
хорошо согласующиеся с экспериментальными и решеточными данными,
дают еще одно свидетельство в пользу предпочтительности значения
параметра нелокальности \myMath{\lambda_q^2=0.4~\gev{2}}
в гауссовых моделях нелокального вакуума КХД.

\section{Подход локальной дуальности}
 \label{sec:LD-approach}
 Правила сумм в подходе ЛД совсем не имеют конденсатных вкладов
из-за предельного перехода \myMath{M^2\rightarrow \infty}.
По этой причине определение порога \myMath{s_0} в таком подходе,
строго говоря, невозможно --- нужно привлекать дополнительные аргументы.
Обычно пользуются тождеством Уорда,
связывающим трехточечный \myMath{AVA}-коррелятор
с двухточечным \myMath{AA}-коррелятором в пределе \myMath{Q^2\to0},
что говорит о точном равенстве порогов
для пионной константы распада \myMath{f_\pi} и для ФФ \myMath{F_\pi(Q^2)}
в подходах ЛД при \myMath{Q^2\to0}.
Это означает,
что при малых значениях \myMath{Q^2\ll1~\gev{2}}
порог \myMath{s_0^{F_\pi(Q^2)}(Q^2)\approx s_0^{f_\pi}}.

Как было показано в работах
\cite{NR82,IS82,NR84-rus,IL84,Rad90,BR91},
основной вклад в ФФ пиона при малых
\myMath{Q^2 \leq 10~\text{ГэВ}^2}
обеспечивается механизмом Фейнмана без обмена жесткими глюонами,
который по этой причине часто называется
\<мягким\> вкладом.
В подходе ЛД этот вклад полностью генерируется
пертурбативной спектральной плотностью~\cite{NR82,Rad95}:
в \myMath{(l+1)}-петлевом порядке\footnote{%
Треугольная диаграмма на однопетлевом уровне с самого начала
не содержит никаких радиационных поправок.
Поэтому \myMath{O(\alpha_s^l)}-поправки для этой диаграммы
возникают в \myMath{(l+1)}-петлевом порядке.}
мы имеем
\begin{subequations}
 \label{eq:FF.LD}
  \begin{eqnarray}
   F_{\pi}^{\text{LD};(l)}(Q^2)
    \!&\!=\!&\! F_{\pi}^{\text{LD};(l)}(Q^2,s_0^{\text{LD};(l)}(Q^2))\, ,\\
   F_{\pi}^{\text{LD};(l)}(Q^2,S)
    \!&\!\equiv\!&\! \frac{1}{f_{\pi}^2}
      \int\limits_{0}^{S}\!\!\!\int\limits_{0}^{S}\!
       \rho_3^{(l)}(s_1, s_2, Q^2)\,ds_1\,ds_2\,,
\end{eqnarray}
\end{subequations}
где \myMath{s_0^{\text{LD};(l)}(Q^2)} ---
эффективный порог ЛД включения высших состояний
в аксиальном канале,
а \myMath{\rho_{3}^{(l)}(s_1, s_2, Q^2)} ---
трехточечная \myMath{(l+1)}-петлевая спектральная плотность.
В ведущем порядке,
\myMath{\rho_{3}^{(0)}(s_1, s_2, Q^2)} известна, см. (\ref{eq:RoSq}),
так что
\begin{eqnarray}
 F_{\pi}^{\text{LD};(0)}(Q^2,S)
  \!&\!=\!&\! \frac{S}{4\pi^2f_\pi^2}\,
       \left[1 - \frac{Q^2 + 6 S}
                      {Q^2 + 4 S}\,
                  \sqrt{\frac{Q^2}{Q^2 + 4 S}}\,
       \right]\,.
\label{eq:FF.LD.LO}
\end{eqnarray}
Предписание ЛД для двухточечного коррелятора
дает соотношения \cite{SVZ,Rad95}
\begin{eqnarray}
 s_0^{\text{LD};(0)}(0)
  = 4\,\pi^2\,f_{\pi}^2
 \quad\text{~и~}\quad
 s_0^{\text{LD};(1)}(0)
  = \frac{4\,\pi^2\,f_{\pi}^2}
         {1+\alpha_s(Q_{0}^2)/\pi}\,,
 \label{eq:LD.s0}
\end{eqnarray}
где \myMath{Q_{0}^2\sim s_0^{\text{LD};(0)}(0)}.
Это предписание есть строгое следствие тождества Уорда
для \myMath{AAV}-коррелятора
из-за сохранения векторного тока.
В принципе, \myMath{Q^2}-зависимость параметра
\myMath{s_0^{\text{LD}}(Q^2)} (\ref{eq:FF.LD})
должна определяться из ПС КХД при \myMath{Q^2\gtrsim1~\gev{2}}.
Однако, как мы уже говорили в разделе~\ref{sec:QCDSR.FF},
стандартный подход ПС КХД становится неприменимым
при \myMath{Q^2>3~\gev{2}}
из-за появления конденсатных вкладов,
линейно растущих по \myMath{Q^2}
\cite{IS83,NR84-rus}.
Поэтому \myMath{Q^2}-зависимость эффективного порога ЛД
была известна только для \myMath{Q^2\lesssim2~\gev{2}},
и по этой причине обычно применялась аппроксимация в виде константы
\cite{NR82,BRS00,BPSS04,BO04},
\myMath{s_0^{\text{LD};(0)}(Q^2)\simeq s_0^{\text{LD};(0)}(0)},
или слабо зависящее от \myMath{Q^2} приближение
\begin{eqnarray}
 \label{eq:LD.s0.(1)}
  s_0^{\text{LD};(1)}(Q^2)
   =
    \frac{4\,\pi^2\,f_{\pi}^2}{1+\alpha_s(Q^2)/\pi}\,,
\end{eqnarray}
как это было сделано в~\cite{BLM07}.

В работе~\cite{BPSS04} мы предложили модель полного ФФ пиона,
основанную на знании факторизуемой части ФФ,
\myMath{F_{\pi}^{\text{pQCD},(2)}(Q^2)},
которая рассчитывалась в коллинеарном приближении
в \myMath{O(\alpha_s^2)}-порядке теории возмущений КХД.
Следует отметить, что факторизуемый пертурбативный вклад
имеет неправильное поведение при \myMath{Q^2=0},
которое должно быть исправлено для выполнения
тождества Уорда (WI) \myMath{F_{\pi}(0)=1}.
Для этого в~\cite{BPSS04}
мы предложили следующую процедуру сшивания:
\begin{eqnarray}
 \label{eq:Fpi-Mod.NNLO}
  F_{\pi}^{\text{WI};(2)}(Q^{2})
   \!&\!=\!&\! F_{\pi}^{\text{LD},(0)}(Q^{2})
     + \left(\frac{Q^2}{2s_0^{(2)}+Q^2}\right)^2
        F_{\pi}^{\text{pQCD},(2)}(Q^2)
\end{eqnarray}
с \myMath{s_0^{(2)}\simeq0.6~\gev{2}}.
Это приближение было использовано для \<склеивания\>
модели ЛД ведущего порядка для \<мягкой\> части,
\myMath{F_{\pi}^{\text{LD},(0)}(Q^{2})},
которая доминирует при малых \myMath{Q^2\leq 1~\gev{2}},
и пертурбативной части жесткого перерассеяния,
\myMath{F_{\pi}^{\text{pQCD},(2)}(Q^2)},
которая учитывает коллинеарные \myMath{O(\alpha_s)+O(\alpha_s^2)}-поправки
и преобладает при больших \myMath{Q^2\gg1~\gev{2}}),
такого чтобы выполнялось тождество Уорда
\myMath{F_{\pi}^{\text{WI};(2)}(0)=1}.
Для проверки качества этой процедуры сшивания
мы проводим сравнение получаемого с ее помощью ФФ
\myMath{F_{\pi}^{\text{WI};(2)}(Q^{2})}
с моделью ЛД (\ref{eq:FF.LD}) на двухпетлевом уровне
(т.~е. в \myMath{O(\alpha_s)}-приближении~\cite{BO04}).
С этой целью мы строим аналогичную \myMath{O(\alpha_s)}-модель
\begin{subequations}
\begin{eqnarray}
 \label{eq:Fpi-Mod.NLO}
  F_{\pi}^{\text{WI};(1)}(Q^{2})
  \!&\!=\!&\!  F_{\pi}^{\text{LD},(0)}(Q^{2})
  + \frac{\alpha_s(Q^2)}{\pi}
      \frac{2\,Q^2\,s_0^{\text{LD};(0)}(0)}
      {(2\,s_0^{\text{LD};(1)}(Q^2)+Q^2)^2}\,,
\end{eqnarray}
где мы использовали выражение для асимптотики пионного ФФ
в пертурбативной КХД \cite{CZS77-rus,FJ79}
\begin{eqnarray}
 \label{eq:Fpi-Pert.NLO}
  F^{\text{pQCD},(1)}_\pi(Q^2)
  = \frac{8\,\pi\,f_\pi^2\,\alpha_s(Q^2)}{Q^2}
  = \frac{\alpha_s(Q^2)}{\pi}\,
        \frac{2\,s_0^{\text{LD};(0)}(0)}{Q^2}\,,
\end{eqnarray}
\end{subequations}
и применяя тот же самый эффективный порог ЛД,
что и в~\cite{BLM07}, а именно,
\myMath{s_0 = s_0^{\text{LD};(1)}(Q^2)}, см.\,(\ref{eq:LD.s0.(1)}).
Следует отметить,
что модель \myMath{F_{\pi}^{\text{WI};(1)}(Q^{2})},
следующая из процедуры сшивания (\ref{eq:Fpi-Mod.NLO}),
работает совсем неплохо,
хотя и была предложена в~\cite{BPSS04} без знания точной
двухпетлевой спектральной плотности,
рассчитанной позже в~\cite{BO04}.
Напомним, что ключевым моментом этой процедуры сшивания
является использование информации о поведении
\myMath{F_{\pi}(Q^2)}
в двух асимптотических режимах
\begin{enumerate}
 \item \myMath{Q^2\to0},
   где тождество Уорда диктует \myMath{F_{\pi}(0)=1}
   и, следовательно,
   \myMath{F_{\pi}(Q^2)\simeq F_{\pi}^{\text{LD},(0)}(Q^2)},
 \item \myMath{Q^2\to\infty},
   где \myMath{F_{\pi}(Q^2)\simeq F_{\pi}^{\text{pQCD},(1)}(Q^2)}
\end{enumerate}
для того, чтобы согласованно объединить \<жесткую\> часть ФФ пиона
с его \<мягкой\> частью.
Численный анализ формулы (\ref{eq:Fpi-Mod.NLO})
показывает, что предложенная процедура сшивания дает достаточно
точный ответ:
относительная ошибка меняется от 5\% при \myMath{Q^2=1~\gev{2}}
до 9\% при \myMath{Q^2=3-30~\gev{2}}.
\begin{figure}[t!]
 \centerline{\includegraphics[width=0.47\textwidth]{
  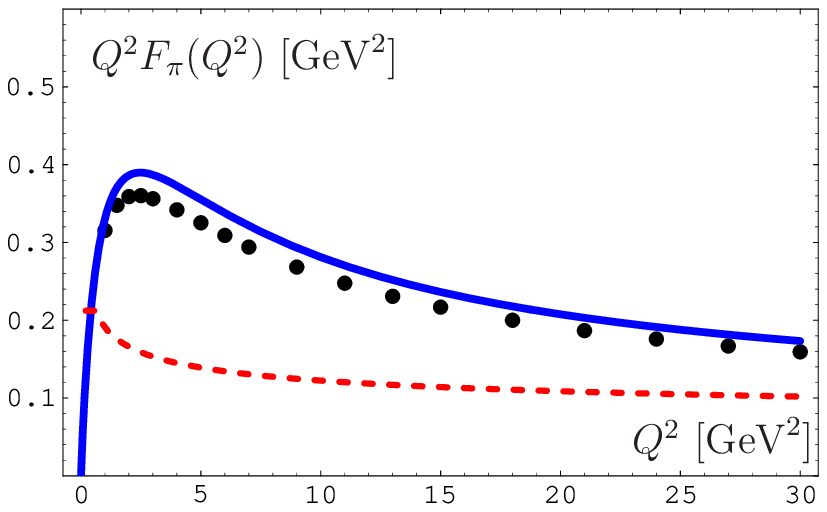}~\includegraphics[width=0.47\textwidth]{
  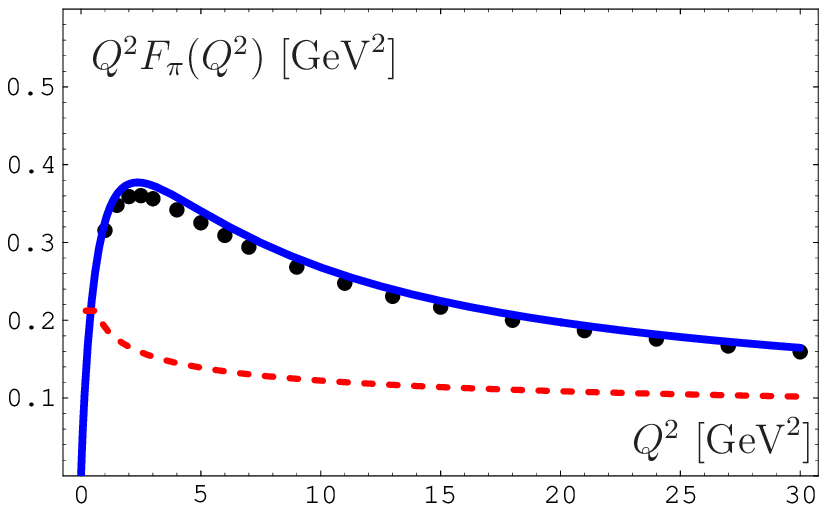}}%
   \caption{Сравнение точного \myMath{O(\alpha_s)}-результата для пионного ФФ
    в подходе локальной дуальности
    \myMath{F_{\pi}^{\text{LD},(1)}(Q^{2})} (черные точки) с моделями
    \myMath{F_{\pi}^{\text{WI};(1)}Q^{2})}  ((\ref{eq:Fpi-Mod.NLO}), слева) и
    \myMath{F_{\pi;\text{imp}}^{\text{WI};(1)}(Q^{2},s_0^{\text{LD};(1)}(Q^2))}
    ((\ref{eq:Fpi-Mod.imp}), справа), показанными на обеих панелях
    сплошными линиями. Асимптотическое пертурбативное КХД-предсказание
    \myMath{F_{\pi}^{\text{pQCD},(1)}(Q^{2})} также показано в виде пунктирной
    линии.\label{fig:Q2FFLDvsMOD}}
\end{figure}
Графическое сравнение \myMath{F_{\pi}^{\text{WI};(1)}(Q^{2})}
(синяя сплошная линия)
с точным результатом ЛД (\ref{eq:FF.LD})
с двухпетлевой спектральной плотностью~\cite{BO04}
(черные точки)
показано на левой панели Рис.\,\ref{fig:Q2FFLDvsMOD}.
Здесь также показан чисто пертурбативный вклад
\myMath{Q^2F_{\pi}^{\text{pQCD},(1)}(Q^{2})}
(красная штрихованная линия) без поправочного фактора согласования
\myMath{[Q^2/(2s_0+Q^2)]^2},
исправляющего поведение в области малых \myMath{Q^2}
(из-за отсутствия этого фактора эта кривая стремится к конечному значению
 0.21~\myMath{\gev{2}} при \myMath{Q^2\to0} и не обращается в 0).

Как только спектральная плотность \myMath{\rho_{3}^{(1)}(s_1, s_2, Q^2)}
стала известной~\cite{BO04},
стало возможным улучшить представление \<мягкой\> части,
описывающей вклад ЛД в (\ref{eq:Fpi-Mod.NLO}),
за счет учета ведущей \myMath{O(\alpha_s)}-поправки
в электромагнитной вершине.
В результате мы уточнили свою прежнюю модель для сшивания
\<мягкой\> и \<жесткой\> частей ---
теперь она должна браться в виде
\myMath{F_{\pi;\text{imp}}^{\text{WI};(1)}(Q^{2},s_0^{\text{LD};(1)}(Q^2))},
где
\begin{eqnarray}
  F_{\pi;\text{imp}}^{\text{WI};(1)}(Q^{2},S)
   \!&\!=\!&\! F_{\pi}^{\text{LD};(0)}(Q^{2},S)
    + \frac{S}{4\pi^2f_\pi^2}\,
        \bigg\{\frac{\alpha_s(Q^2)}{\pi}\,
                \left(\frac{2S}{2S+Q^2}\right)^2
   \nonumber\\
   \!&\!\!&~~~~~~~~~~~~~~~~~~~~~~~~
   +\,F^{\text{pQCD},(1)}_\pi(Q^2)\,
               \left(\frac{Q^2}{2S+Q^2}\right)^2
        \bigg\}\,.~~~~~~
 \label{eq:Fpi-Mod.imp}
\end{eqnarray}
Здесь мы явно ввели зависимость от порога \myMath{S},
чтобы применять эту формулу позднее
с порогом \myMath{S=s_0^{\text{LD-eff}}(Q^2)},
извлеченным из сравнения полученных нами результатов
в ПС КХД с НВК с \myMath{O(\alpha_s)}-подходом ЛД.
Отметим, что предложенное нами улучшение
имеет правильное поведение при \myMath{Q^2\to0},
если при этом
\myMath{S\simeq s_0^{\text{LD};(1)}(Q^2)=4\pi^2f_\pi^2/(1+\alpha_s(Q^2)/\pi)}.
В самом деле, мы имеем
\myMath{F_{\pi}^{\text{LD};(0)}(0,S)=S/(4\pi^2f_\pi^2)}
и за счет того,
что вклад \myMath{F^{\text{pQCD},(1)}_\pi(Q^2)}
в этом пределе
обращается в ноль благодаря фактору \myMath{[Q^2/(2S+Q^2)]^2},
полный результат есть\footnote{%
Обычно применяется \<замораживание\> порога
\myMath{s_0^{\text{LD};(1)}(Q^2) = 4\pi^2f_\pi^2/(1+\alpha_s(s_0)/\pi)} при
\myMath{Q^2\leq s_0\simeq0.6~\gev{2}}.
В этом случае такое же \<замораживание\> должно использоваться
в (\ref{eq:Fpi-Mod.imp}) для аргумента \myMath{\alpha_s}:
\myMath{\alpha_s(Q^2)\to\alpha_s(s_0)} при \myMath{Q^2\leq s_0}.
Напомним, что для правильного расчета электромагнитного радиуса пиона
(т.~e. производной пионного ФФ в области малых \myMath{Q^2})
необходимо применять другой тип операторного разложения
и соответствующее ему другое ПС --- детали см. в~\cite{NR84-rus}.}
$$F_{\pi;\text{imp}}^{\text{WI};(1)}(0,s_0^{\text{LD};(1)}(0))
= \frac{s_0^{\text{LD};(1)}(0)}{4\pi^2f_\pi^2}
  \left[1+\frac{\alpha_s(s_0)}{\pi}\right]
= 1.
$$
Графическое сравнение точности улучшенной модели в сравнении
с точным ответом ЛД в однопетлевом приближении
проведено на правой панели Рис.\,\ref{fig:Q2FFLDvsMOD}.
Хорошо видно, что новая формула сшивания (\ref{eq:Fpi-Mod.imp})
работает лучше: ее относительная ошибка уменьшилась
до 4\% при \myMath{Q^2=1-10~\gev{2}}
и 3\% при \myMath{Q^2=30~\gev{2}}.

Продолжая предложенную процедуру на двухпетлевой уровень,
мы получаем двухпетлевую оценку пионного ФФ,
\myMath{F_{\pi}^{\text{WI};(2)}(Q^{2},s_0^{\text{LD};(2)}(Q^2))},
в следующем виде:
\begin{eqnarray}
  F_{\pi}^{\text{WI};(2)}(Q^{2},S)
   \!&\!=\!&\! F_{\pi}^{\text{LD};(0)}(Q^{2},S)
     + \frac{S}{4\pi^2f_\pi^2}\,
        \bigg\{\frac{\alpha_s(Q^2)}{\pi}\,
                \left(\frac{2S}{2S+Q^2}\right)^2
   \nonumber\\
  \!&\!\!&\!~~~~~~~~~~~~~~~~~~~~~~~~~~
            + F^{\text{FAPT},(2)}_\pi(Q^2)\,
               \left(\frac{Q^2}{2S+Q^2}\right)^2
        \bigg\}\,,~~~~~~~
 \label{eq:Fpi.WI-Mod.NNLO}
\end{eqnarray}
где \myMath{F^{\text{FAPT},(2)}_\pi(Q^2)} есть
аналитическое выражение,
которое получено из \myMath{F^{\text{pQCD},(2)}_\pi(Q^2)}
с помощью ДАТВ
при выборе не зависящего от \myMath{Q^2} значения
масштаба факторизации \myMath{\muF}:
\begin{eqnarray}
  F_{\pi}^{\text{FAPT},(2)}(Q^2)
   \!&\!=\!&\! \mathcal A_{1}^{(2)}(Q^2)\,
        \mathcal F_{\pi}^\text{LO}(Q^2;\muF)
     + \frac{1}{\pi}\,
        \mathcal L_{2;1}^{(2)}(Q^2)\,
         \mathcal F_{\pi}^{(1,\text{F})}(Q^2;\muF)
  \nonumber\\
 \label{eq:piff.NNLO.APT}
   \!&\!+\!&\! \frac{1}{\pi}\,
        \mathcal A_{2}^{(2)}(Q^2)\,
         \left[\mathcal F_{\pi}^\text{NLO}(Q^2;\muF)
             - \mathcal F_{\pi}^{(1,\text{F})}(Q^2;\muF)\,
                \ln\frac{Q^2}{\Lambda_{3}^2}
         \right]~~~~~~~
\end{eqnarray}
и которое дает для факторизованной части ФФ результат,
очень близкий к получаемому в АТВ при стандартном выборе масштабов,
\myMath{\muR=\muF=Q^2}.

Эта модель дает нам возможность применить \myMath{O(\alpha_s^2)}-результаты
коллинеарного приближения пертурбативной КХД
для расчета ФФ пиона в двухпетлевом приближении
без явного счета трехпетлевой трехточечной спектральной плотности
--- а этот счет достаточно сложен.
Отметим, что при моделировании однопетлевого выражения ЛД
в этом подходе с помощью (\ref{eq:Fpi-Mod.imp}),
как мы только что показали
(см. левую панель Рис.\,\ref{fig:Q2FFLDvsMOD}),
ошибка получается не более 10\%.
Поскольку относительный вклад двухпетлевой поправки в ФФ пиона
сам по себе имеет порядок 10\%, см.~\cite{BPSS04,AB08rus},
то относительная ошибка нашей оценки
\myMath{O(\alpha_s^2)}-поправки получается на уровне 1\%
--- если только мы учли \myMath{O(\alpha_s)}-поправку
точно за счет специального выбора \myMath{s_0(Q^2)},
как это и делается в (\ref{eq:FF.LD.s0}).
То есть все, что нам надо сейчас сделать,
это построить эффективные пороги ЛД
\myMath{s_0^{\text{LD};(1)}(Q^2)}.

\begin{figure}[t!]
 \centerline{\includegraphics[width=0.45\textwidth]{
  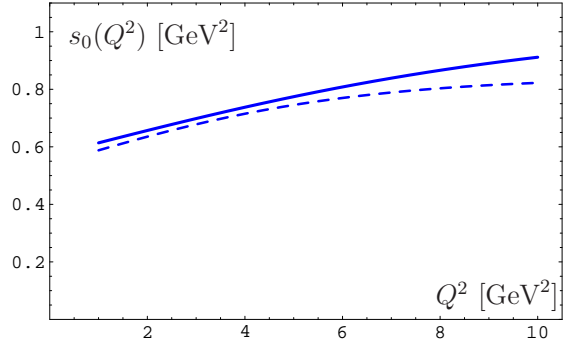}}
   \caption{Эффективные пороги континуума
     \myMath{s_{0,\text{imp}}^\text{LD-eff}(Q^2)} (сплошная линия) и
     \myMath{s_{0,\text{min}}^\text{LD-eff}(Q^2)} (штрихованная линия),
     аппроксимирующие результаты ПС КХД с НВК с помощью
     \myMath{O(\alpha_s)}-формул локальной дуальности (\ref{eq:Fpi-Mod.imp}).
     \label{fig:s0LD}}
\end{figure}
Как мы уже говорили в самом начале этого раздела,
задача получения эффективного порога континуума
\myMath{s_0^{\text{LD}}(Q^2)} является центральной
для подхода ЛД.
С нашей точки зрения,
они должны определяться из условия совпадения
результатов ЛД с результатами борелевских ПС КХД.
В предыдущем разделе мы построили и обработали такие ПС
и получили интерполяционные формулы
(\ref{eq:FF.pi.SR.Int.Min})--(\ref{eq:FF.pi.SR.Int.Imp}),
которые применимы для
\myMath{Q^2\in[1,10]~\gev{2}}.
Теперь с их помощью мы можем определить соответствующие
эффективные пороги \myMath{s_{0,\text{min}}^\text{LD-eff}(Q^2)}
и \myMath{s_{0,\text{imp}}^\text{LD-eff}(Q^2)}
в минимальной и улучшенной гауссовых моделях вакуума КХД,
соответственно:
\begin{subequations}
 \label{eq:FF.LD.s0}
\begin{eqnarray}
  F_{\pi;\text{imp}}^{\text{WI};(1)}
  \left(Q^2,s_{0,\text{min}}^\text{LD-eff}(Q^2)\right)
  \!&\!=\!&\! F_{\pi;\text{min}}^{\text{SR}}(Q^2)\,,
  \\
  F_{\pi;\text{imp}}^{\text{WI};(1)}
  \left(Q^2,s_{0,\text{imp}}^\text{LD-eff}(Q^2)\right)
  \!&\!=\!&\! F_{\pi;\text{imp}}^{\text{SR}}(Q^2)\,.~~~~
\end{eqnarray}
\end{subequations}
\begin{figure}[b!]
 \centerline{\includegraphics[width=0.47\textwidth]{
  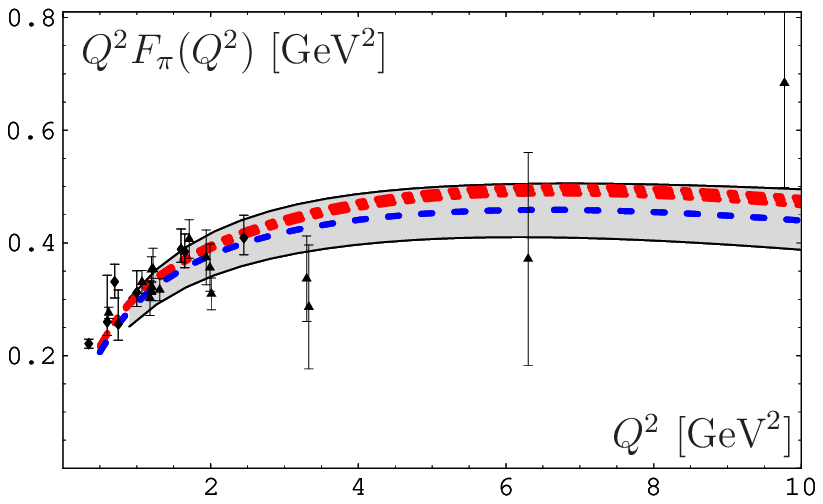}~\includegraphics[width=0.47\textwidth]{
  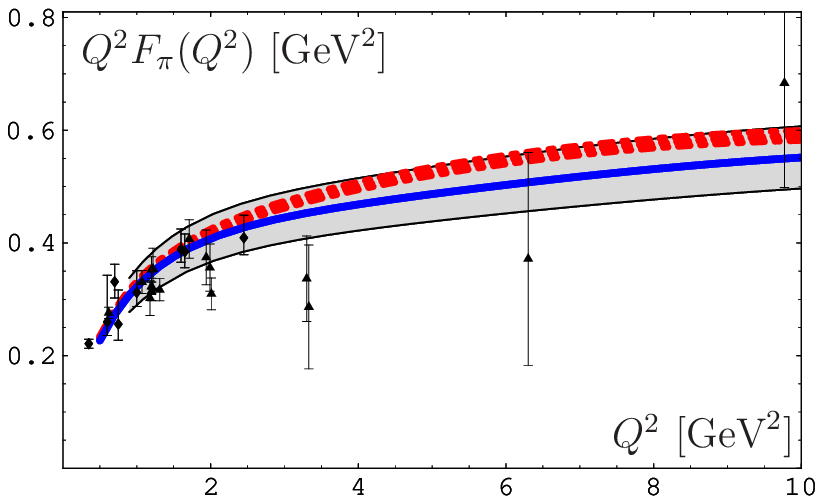}}%
     \caption{Сравнение предсказаний для пионного ФФ \myMath{Q^2 F_{\pi}(Q^2)},
   полученных по формуле (\ref{eq:Fpi.WI-Mod.NNLO}) с использованием
   эффективных порогов континуума \myMath{s_{0,\text{min}}^\text{LD-eff}(Q^2)}
   (слева) и \myMath{s_{0,\text{imp}}^\text{LD-eff}(Q^2)} (справа).
   Результаты ПС КХД с НВК показаны на обеих панелях теми же затененными
   полосами, что и на Рис.~\ref{fig:main-result}, в то время как
   \myMath{O(\alpha_s^2)}-предсказаниям отвечают жирные штрихованные полосы,
   слегка превышающие центральные линии затененных полос.\label{fig:FF.LD.2Loop}}
\end{figure}
Решения этих уравнений, а именно,
\myMath{s_{0,\text{min}}^\text{LD-eff}(Q^2)}
и \myMath{s_{0,\text{imp}}^\text{LD-eff}(Q^2)},
показаны на Рис.\,\ref{fig:s0LD}.
Они могут быть представлены в этой области значений
\myMath{Q^2}
такими интерполяционными формулами:
\begin{subequations}
 \label{eq:FF.LD.s0.App}
\begin{eqnarray}
  s_{0,\text{min}}^\text{LD-eff}(Q^2=x~\text{ГэВ}^2)
  \!&\!=\!&\! 0.57+0.307\,\tanh(0.165\,x)
            - 0.0323\,\tanh(775\,x)\, ,\\
  s_{0,\text{imp}}^\text{LD-eff}(Q^2=x~\text{ГэВ}^2)
  \!&\!=\!&\! 0.57+0.461\,\tanh(0.0954\,x)\,.~~~
\end{eqnarray}
\end{subequations}
Мы видим,
что оба порога умеренно растут с ростом \myMath{Q^2}.

Результаты для пионного ФФ, полученные с помощью нашей
двухпетлевой модели (\ref{eq:Fpi.WI-Mod.NNLO})
и эффективных порогов континуума
\myMath{s_{0,\text{min}}^\text{LD-eff}(Q^2)} и
\myMath{s_{0,\text{imp}}^\text{LD-eff}(Q^2)},
показаны на Рис.\,\ref{fig:FF.LD.2Loop}.
Конечная ширина штрихованных полос связана с учетом разброса
параметров \myMath{a_2} и \myMath{a_4} пионной АР в пучке моделей,
диктуемых ПС КХД с НВК~\cite{BMS01,BMS01c,BP06rus}.
Из этого рисунка мы видим,
что самое большое значение двухпетлевая поправка дает в области
\myMath{Q^2\gtrsim4~\gev{2}},
достигая здесь уровня \myMath{3-10}\%,
что на количественно неплохо совпадает с оценками,
полученными нами ранее в ДАТВ~\cite{BPSS04,AB08rus}.

\section{Заключение}
 \label{sec:Conclusions}
 Мы построили и проанализировали для двух гауссовых моделей
нелокального КХД-вакуума трехточечное правило сумм КХД
для электромагнитного ФФ пиона.
К достоинствам такого анализа следует отнести
независимость результатов от профиля пионной АР ---
это приводит
к уменьшению теоретической неопределенности,
связанной с параметризацией пионной АР
в низкой точке нормировки (порядка 1~ГэВ).
Кроме того, мы попытались уменьшить влияние нарушения калибровочной инвариантности,
индуцируемой грубым моделированием нелокальности вакуумных конденсатов:
в наших расчетах мы учли не только минимальную гауссову модель НВК,
но также и улучшенную модель, разработанную нами в~\cite{BP06rus}.

Полученные нами результаты можно суммировать следующим
перечнем.
\begin{enumerate}
  \item
  Основные предсказания для ФФ пиона \myMath{F_\pi(Q^2)} показаны
  на Рис.\ \ref{fig:main-result} в сравнении с существующими
  экспериментальными данными (старыми) Корнелла \cite{FFPI74,FFPI76,FFPI78}
  и (более новыми) группы JLab \cite{JLab08II}.
  Оказалось, что учет \myMath{O(\alpha_s)}-вклада в спектральную плотность
  повысил предсказание для ФФ в среднем на 20\%,
  что несколько меньше предыдущих оценок относительной важности
  таких поправок \cite{BO04,BLM07}.
  \item
  Как хорошо видно из Рис.\ \ref{fig:main-result} центральная линия
  предсказаний для ФФ пиона, полученных в улучшенной модели,
  лежит внутри полосы ошибок минимальной модели вплоть до значения
  \myMath{Q^2\approx 7~\gev{2}},
  что говорит о сравнимом качестве предсказаний обеих гауссовых моделей НВК
  в этой области.
  Основное отличие между этими двумя моделями связано с различными
  вкладами кварк-глюон-антикваркового конденсата
  \myMath{\Phi_{\bar qAq}(Q^2,M^2)} (см. Рис.\ \ref{fig:Q2FFapart}).
  \item
  Полученные нами предсказания хорошо согласуются
  с результатами недавних решеточных расчетов~\cite{Brommel06}
  пионного ФФ при \myMath{Q^2\lesssim4~\gev{2}}.
  \item
  Мы построили также эффективные пороги континуума
  для подхода локальной дуальности,
  которые позволяют в этом подходе имитировать
  результаты, полученные нами в подходе борелевских ПС КХД.
  Сравнение наших предсказаний с
  полученными ранее в подходе локальной дуальности \cite{BLM07}
  выявляет систематическое превышение наших результатов.
  Причиной такой разницы является тот факт,
  что эффективный порог континуума
  \myMath{s_0^\text{LD}(Q^2)}
  хорошо определен только в области малых \myMath{Q^2}.
  Для б\'{о}льших \myMath{Q^2} авторы~\cite{BLM07}
  предложили пользоваться логарифмически растущим порогом
  \begin{eqnarray*}
    s_0^\text{LD}(Q^2)
    =
   \frac{4\pi^2 f_\pi^2}{1+\alpha_s(Q^2)/\pi}\,,
  \end{eqnarray*}
  равным \myMath{0.67~\gev{2}} при \myMath{Q^2\approx10~\gev{2}}.
  Построенный нами эффективный порог континуума,
  имитирующий результат ПС КХД, оказывается
  выше: \myMath{s_0^\text{LD}(Q^2=10~\text{ГэВ}^2)=0.87~\text{ГэВ}^2}.
  Это означает, что ошибка в определении \myMath{s_0^\text{LD}}
  в области \myMath{Q^2=10~\gev{2}} оказывается порядка \myMath{20}\%.
  \item
  На основе тождества Уорда мы построили \cite{BPSS04,BPS09} процедуру согласования
  факторизуемой части ФФ пиона,
  рассчитываемой в коллинеарной КХД,
  с моделью нефакторизуемого вклада,
  получаемой в подходе локальной дуальности
  с использованием построенных эффективных порогов континуума.
  С ее помощью мы построили выражение для ФФ пиона
  в \myMath{O(\alpha_s^2)}-приближении в развитом нами подходе ДАТВ.
\end{enumerate}

\section*{Благодарности}
  Эта работа была выполнена при финансовой поддержке грантов РФФИ
№~08-01-00686 и 09-02-01149,
программы сотрудничества БРФФИ--ОИЯИ (контракт №~F10D-001),
гранта 2010~г. программы Гейзенберг--Ландау.

\newcommand{\myAppSect}[1]{\refstepcounter{section}\setcounter{figure}{0}
 \setcounter{table}{0}\setcounter{equation}{0}
  \section*{Приложение \arabic{section}.~#1}}
   \renewcommand {\theequation}{П\arabic{section}.\arabic{equation}}
    \renewcommand {\thefigure}{П\arabic{section}.\arabic{figure}}
     \renewcommand {\thesection}{\arabic{section}}
      \setcounter{section}{0}
 \myAppSect{Параметризация нелокальных вакуумных конденсатов}
  \label{App:Param.NLC}
Как принято в подходе ПС КХД,
мы используем калибровку Фока--Швингера
$x^\mu A_\mu(x)=0$.
По этой причине все струнные вставки
${\mathcal C}(x,0)\equiv
 {\mathcal P}\exp\!\left[-ig_s\!\!\int_0^x t^{a} A_\mu^{a}(y)dy^\mu\right]=1$,
необходимые для обеспечения калибровочной инвариантности
нелокальных объектов,
обращаются в единицу
при выборе контура интегрирования в виде прямой линии,
соединяющей точки $0$ и $x$.

Минимальная гауссова модель нелокального вакуума КХД
отвечает следующему  анзацу:
\begin{eqnarray}
 \label{eq:Min.Anz.qGq}
  f_i\left(\alpha,\beta,\gamma\right)
   = \delta\left(\alpha -\Lambda\right)\,
     \delta\left(\beta  -\Lambda\right)\,
     \delta\left(\gamma -\Lambda\right)
\end{eqnarray}
с параметром $\Lambda=\lambda_q^2/2$.
Как было показано в наших работах~\cite{BP06rus,BP06zako},
в этой модели нарушены уравнения движения КХД
и двухточечный коррелятор векторных токов
имеет продольную составляющую.
Для того, чтобы выполнить уравнение движения КХД,
связывающее векторные билокальный (\ref{eq:V.NLC})
и кварк-глюон-антикварковые конденсаты (\ref{eq:qAq.mu.nu}),
и в то же время минимизировать непоперечность $VV$-коррелятора,
нами была предложена улучшенная гауссова модель
вакуума КХД~\cite{BP06rus}:
\begin{subequations}
\label{eq:Imp.Anz.qGq}
\begin{eqnarray}
 \label{eq:Imp.Anz.qGq.f}
  f^\text{imp}_i\left(\alpha,\beta,\gamma\right)
   = \left(1 + X_{i}\partial_{x}
            + Y_{i}\partial_{y}
            + Y_{i}\partial_{z}
    \right)
         \delta\left(\alpha-x\Lambda\right)
         \delta\left(\beta -y\Lambda\right)
         \delta\left(\gamma-z\Lambda\right)\,,~~~
\end{eqnarray}
где $z=y$, $\Lambda=\frac12\lambda_q^2$ и
\begin{eqnarray}
  X_1 \!&\!=\!&\! +0.082\,;~X_2 = -1.298\,;~X_3 = +1.775\,;~x=0.788\, ,~~~\\
  Y_1 \!&\!=\!&\! -2.243\,;~Y_2 = -0.239\,;~Y_3 = -3.166\,;~y=0.212\,.~~~
\end{eqnarray}
\end{subequations}
За счет уравнения движения КХД эти параметры удовлетворяют
следующим условиям:
\begin{eqnarray}
 \label{eq:3L.D.A.Rule4}
  12\,\left(X_{2} + Y_{2}\right)
  - 9\,\left(X_{1} + Y_{1}\right)
  = 1 \,,~~~~~x+y=1\,.
\end{eqnarray}

Нелокальные вакуумные конденсаты 4-кварковых операторов,
как обычно,
сведены к произведениям двух скалярных кварковых конденсатов
(\ref{eq:S.NLC})
с использованием гипотезы вакуумной доминантности \cite{SVZ}.

 \myAppSect{Численные параметры для правил сумм КХД}
  \label{App:QCDSR.Param}
Мы приводим здесь выражения для коэффициентных функций
\myMath{P_i(M^2,a,b,x,y)} и
\myMath{S_i(M^2,a,b,x,y)},
которые определяют вклад кварк-глюон-антикварковых конденсатов,
\myMath{\Phi_{\bar{q}Aq}(M^2,Q^2)},
в ПС для ФФ пиона, см.~(\ref{eq:Phi.qAq}):\footnote{
Для сокращения записи аргументы этих функций не выписаны ---
предполагается, что они имеют везде один и тот же вид,
именно, \myMath{P_i(M^2,a,b,x,y)} и \myMath{S_i(M^2,a,b,x,y)}.}
Сначала выпишем коэффициентные функции
\myMath{P_i(M^2,a,b,x,y)}:
\begin{subequations}
 \label{eq:qAq.SR.Pi}
\begin{eqnarray}
 \label{eq:qAq.SR.P1}
  P_1
   \!&\!=\!&\!
     \bar{y}\,
      \left[\frac{3-a}{2}\right]
   + \frac{\bar{y}}{\bar{x}}\,
      \left[\frac{a-5}{2}-\frac{b}{\bar{a}}\right]
   + \bar{b}\,
      \left[\frac{a+1}{2}-\frac{1}{1+a}+\delta(a)\right]
  \\
   \!&\!+\!&\!
     \bar{b}\,\bar{x}\,
      \left[\frac{5-a}{2}-\frac{4}{1+a}\right]
   + \bar{b}\,\bar{y}\,
      \left[2a-6+\frac{4}{1+a}+\frac{3}{2}\,\delta(a)\right]
  \nonumber\\
   \!&\!+\!&\!
     \frac{\bar{b}\,\bar{y}}{\bar{x}}\,
      \left[1-2a+\frac{1}{1+a}-\frac{1}{2}\,\delta(a)+\delta(\bar{a})\right]
  \nonumber\\
   \!&\!+\!&\!
     \frac{\bar{b}\,Q^2}{2M^2}\,
      \left\{
        a-7+\frac{20}{1+a}-\frac{16}{(1+a)^2}
     + \frac{1}{\bar{y}}\,
        \left[a-\frac{2}{1+a}+\frac{2}{(1+a)^2}\right]
      \right\}
  \nonumber\\
   \!&\!-\!&\!
     \frac{\bar{b}\,Q^2}{2M^2}\,
      \left\{
       \frac{1}{\bar{x}}\,
        \left[a+3-\frac{10}{1+a}+\frac{4}{(1+a)^2}\right]
     - \frac{\bar{y}}{\bar{x}}\,
        \left[1-\frac{6}{1+a}+\frac{8}{(1+a)^2}\right]
      \right\}
  \nonumber\\
   \!&\!-\!&\!
     \frac{\bar{b}\,Q^2}{2M^2}\,
      \left\{
       \frac{\bar{x}}{\bar{y}}\,
        \left[a-6+\frac{14}{1+a}-\frac{8}{(1+a)^2}\right]
     - \frac{\bar{y}}{\bar{x}^2}\,
        \left[3-\frac{8}{1+a}+\frac{2}{(1+a)^2}\right]
      \right\};
  \nonumber\\ \nonumber\\
 \label{eq:qAq.SR.P2}
  P_2
   \!&\!=\!&\!
    \frac{-2\,\bar{y}\,\bar{a}}{\,\bar{x}}\,;\\
 \label{eq:qAq.SR.P3}
  P_3
   \!&\!=\!&\! P_1
  + \frac{4\,\bar{y}}{\,\bar{x}}\,;
\end{eqnarray}
\end{subequations}

Теперь коэффициентные функции
\myMath{S_i(M^2,a,b,x,y)}:
\begin{subequations}
 \label{eq:qAq.SR.Si}
\begin{eqnarray}
 \label{eq:qAq.SR.S1}
  S_1
   \!\!&\!=\!&\!
     \bar{y}\,
      \left[\frac{3\bar{a}}{2}+\frac{2b}{\bar{a}}\right]
   + \frac{\bar{y}}{\bar{x}}\,
      \left[\frac{3a+7}{2}-\frac{3b}{\bar{a}}\right]
   + \bar{b}\,
      \left[\frac{3a-7}{2}-\frac{1}{1+a}+\delta(a)\right]
  \\
   \!\!&\!+\!&\!
     \bar{b}\,\bar{x}\,
      \left[\frac{9-3a}{2}-\frac{8}{1+a}\right]
   + \bar{b}\,\bar{y}\,
      \left[6a-9+\frac{8}{1+a}+\frac{7}{2}\,\delta(a)\right]
  \nonumber\\
   \!\!&\!+\!&\!
     \frac{\bar{b}\,\bar{y}}{\bar{x}}\,
      \left[\delta(\bar{a})-6a-2+\frac{1}{1+a}-\frac{1}{2}\,\delta(a)\right]
  \nonumber\\
   \!\!&\!+\!&\!
     \frac{\bar{b}\,Q^2}{2M^2}
      \left\{
        3a-15+\frac{40}{1+a}-\frac{32}{(1+a)^2}
     + \frac{1}{\bar{y}}
        \left[3a-10+\frac{8}{1+a}+\frac{2}{(1+a)^2}\right]
      \right\}
  \nonumber\\
   \!\!&\!+\!&\!
     \frac{\bar{b}\,Q^2}{2M^2}
      \left\{
       \frac{1}{\bar{x}}
        \left[9-3a-\frac{2}{1+a}-\frac{4}{(1+a)^2}\right]
     + \frac{\bar{y}}{\bar{x}}
        \left[3-\frac{12}{1+a}+\frac{16}{(1+a)^2}\right]
      \right\}
  \nonumber\\
   \!\!&\!-\!&\!
     \frac{\bar{b}\,Q^2}{2M^2}
      \left\{
       \frac{\bar{x}}{\bar{y}}
        \left[3a-12+\frac{28}{1+a}-\frac{16}{(1+a)^2}\right]
     - \frac{\bar{y}}{\bar{x}^2}
        \left[1-\frac{6}{1+a}+\frac{2}{(1+a)^2}\right]
      \right\}; \nonumber
\end{eqnarray}
\begin{eqnarray}
 \label{eq:qAq.SR.S2}
  S_2
   \!\!&\!=\!&\!
     \bar{y}\,
      \left[a^2+\frac{a-3}{2}+\frac{2\bar{b}}{\bar{a}}\right]
   - \frac{\bar{y}}{\bar{x}}\,
      \left[a^2+\frac{a-5}{2}+\frac{3b}{\bar{a}}\right]
   \\
   \!\!&\!-\!&\!
     \bar{b}\,
      \left[a^2+\frac{a+1}{2}+\frac{1}{1+a}-\delta(a)\right]
   - \bar{b}\,\bar{y}\,
      \left[5a^2-2a-4-\frac{1}{2}\,\delta(a)\right]
  \nonumber\\
   \!\!&\!+\!&\!
     \bar{b}\,\bar{x}\,
      \left[a^2-\frac{3a+1}{2}\right]
   + \frac{\bar{b}\,\bar{y}}{\bar{x}}\,
      \left[\delta(\bar{a})+5a^2-2a+7+\frac{1}{1+a}-\frac{3}{2}\,\delta(a)\right]
  \nonumber\\
   \!\!&\!-\!&\!
     \frac{\bar{b}\,Q^2}{2M^2}
      \left\{
        2a^2-7a+9-\frac{8}{1+a}
   + \frac{1}{\bar{y}}
        \left[2a^2-a+2-\frac{2}{(1+a)^2}\right]
           \right\}
  \nonumber\\
   \!\!&\!+\!&\!
     \frac{\bar{b}\,Q^2}{2M^2}
      \left\{
       \frac{1}{\bar{x}}
        \left[2a^2-7a+11-\frac{2}{1+a}-\frac{4}{(1+a)^2}\right]
     - \frac{\bar{y}}{\bar{x}}
        \left[2a-5+\frac{4}{1+a}\right]
      \right\}
  \nonumber\\
   \!\!&\!+\!&\!
     \frac{\bar{b}\,Q^2}{2M^2}
      \left\{
       \frac{\bar{x}}{\bar{y}}
        \left[2a^2-5a+4-\frac{4}{1+a}\right]
     + \frac{\bar{y}}{\bar{x}^2}
        \left[2a-5+\frac{2}{1+a}+\frac{2}{(1+a)^2}\right]
      \right\}; \nonumber\\ \nonumber\\
 \label{eq:qAq.SR.S3}
  S_3
   \!\!&\!=\!&\! P_1
  + \frac{4\,\bar{y}}{\,\bar{x}}\,;
\end{eqnarray}
\end{subequations}
Отметим, что \<опасные\> знаменатели присутствуют в этих выражениях
только в виде комбинаций \myMath{b/\bar{a}},
которые на самом деле не опасны,
если вспомнить, что в подынтегральных функциях
\myMath{f_i(1/\sigma,1/\tau,1/\rho)}
имеются аргументы вида \myMath{1/\tau\equiv M^2\bar{x}b/\bar{a}}:
при стремлении \myMath{b/\bar{a}\to\infty}
аргумент \myMath{1/\tau} также будет неограниченно возрастать,
а подынтегральные функции \myMath{f_i(1/\sigma,1/\tau,1/\rho)}
очень быстро будут стремиться к нулю.
Поэтому в интеграле по \myMath{\tau},
в который преобразуется интеграл по \myMath{b},
никаких особенностей не остается:
\begin{eqnarray*}
 \int\limits_0^1 \varphi_i(M^2,a,b,x,y)\frac{b\,db}{1-a}
  \!&\!\sim\!&\!
   \frac{1}{\bar{x}(a+1)}
    \int\limits_0^1 f_i(1/\sigma,1/\tau,1/\rho)\frac{d\tau}{\tau^3}
 \stackrel{a\to1}{\rightarrow}
   \frac{1}{2\bar{x}}
    \int\limits_1^{\infty} f_i(1/\sigma,\tau,1/\rho)\tau d\tau\,.
\end{eqnarray*}

Теперь приведем численные значения вакуумных конденсатов:
$\lambda_q^2=0.4$~ГэВ${}^2$,
$\langle\alpha_s{GG}\rangle/\pi=0.012$~ГэВ$^4$
и
$\alpha_s\,\langle\bar{q}q\rangle^2$ = $1.83\cdot 10^{-4}$~ГэВ$^6$.
Нелокальный глюонный конденсат $\Phi_{\va{GG}}(M^2)$
приводит к очень сложному выражению,
которое по аналогии с кварковым случаем
моделируется экспоненциальным фактором
\cite{BR91,MS93-rus}:
$\Phi_{\va{GG}}(M^2) =
 \Phi_{\va{GG}}^\text{loc}(M^2)\,e^{-\lambda_g^2 Q^2/M^4}$
с $\lambda_g^2=0.4$~ГэВ${}^2$.

Для определения наилучшего порога континуума $s_0$
мы для каждого значения $Q^2$ и $s_0$
вводим $\chi^2$-функцию:
\begin{eqnarray}
 \label{eq:xi}
 \chi^2(Q^2,s_0)
 = \frac{\varepsilon^{-2}}{N_M}
    \left[\sum\limits_{i=0}^{N_M}
           Q^4\,F(Q^2,M^2_{i},s_0)^2
        - \frac{\left(\sum\limits_{i=0}^{N_M}
                       Q^2\,F(Q^2,M^2_{i},s_0)
                \right)^2}
               {N_M+1}
    \right]\!,~~~
\end{eqnarray}
где $M_{i}^2=M_{-}^2+i\,\Delta_M$,
$\Delta_M=(M_{+}^2-M_{-}^2)/N_M$,
$N_M=20$,
и $\varepsilon$ обозначает желаемую точность определения ФФ $F(Q^2,M^2_{i},S)$
при $\chi^2\simeq1$
(мы использовали значение $\varepsilon=0.07$~ГэВ${}^2$.)


\end{document}